\documentclass[aps,prb,reprint,superscriptaddress]{revtex4-2}

\usepackage[english]{babel}
\usepackage[utf8x]{inputenc}
\usepackage[T1]{fontenc}
\usepackage{siunitx}
\usepackage{xcolor}
\usepackage{xspace}
\usepackage[version=4]{mhchem}

\usepackage{placeins}
\usepackage{braket}
\usepackage{longtable}
\usepackage{booktabs}
\usepackage{natbib}
\usepackage{bibentry}
\usepackage{amsmath}
\usepackage{amssymb}
\usepackage{graphicx}

\usepackage{ragged2e}

\renewcommand{\deg}{$^{\circ}$}

\begin{document}
%\linenumbers
%\setlength\linenumbersep{5pt}

\title{A silicon spin vacuum: isotopically enriched $^{28}$silicon-on-insulator and $^{28}$silicon from ultra-high fluence ion implantation}

\author{Shao Qi Lim}
\email{qi.lim@unimelb.edu.au}
\affiliation{Centre for Quantum Computation and Communication Technology, School of Physics, The University of Melbourne, Parkville VIC Australia.}
\affiliation{School of Science, RMIT University, Melbourne, VIC Australia.}

\author{Brett C. Johnson}
\affiliation{School of Science, RMIT University, Melbourne, VIC Australia.}

\author{Sergey Rubanov}
\affiliation{Ian Holmes Imaging Centre, Bio21 Institute, The University of Melbourne, Melbourne VIC, Australia.}

\author{Nico Klingner}
\affiliation{Helmholtz-Zentrum Dresden-Rossendorf (HZDR), Institute of Ion Beam Physics and Materials Research, Dresden, Germany}

\author{Bin Gong}
\affiliation{Mark Wainwright Analytical Centre, University of New South Wales Sydney, Sydney NSW, Australia.}

\author{Alexander M. Jakob}
\affiliation{Centre for Quantum Computation and Communication Technology, School of Physics, The University of Melbourne, Parkville VIC Australia.}

\author{Danielle Holmes}
\affiliation{Centre for Quantum Computation and Communication Technology, School of Electrical Engineering and Telecommunications, UNSW Australia, NSW Australia.}

\author{David N. Jamieson}
\affiliation{Centre for Quantum Computation and Communication Technology, School of Physics, The University of Melbourne, Parkville VIC Australia.}

\author{Jim S. Williams}
\affiliation{Research School of Physics, The Australian National University, Canberra ACT, Australia.}

\author{Jeffrey C. McCallum}
\affiliation{Centre for Quantum Computation and Communication Technology, School of Physics, The University of Melbourne, Parkville VIC Australia.}

\date{\today}

\begin{abstract}
Isotopically enriched silicon (Si) can greatly enhance qubit coherence times by minimizing naturally occurring $^{29}$Si which has a non-zero nuclear spin. Ultra-high fluence $^{28}$Si ion implantation of bulk natural Si substrates was recently demonstrated as an attractive technique to ultra-high $^{28}$Si isotopic purity. In this work, we apply this $^{28}$Si enrichment process to produce $^{28}$Si and $^{28}$Si-on-insulator (SOI) samples. Experimentally, we produced a $^{28}$Si sample on natural Si substrate with $^{29}$Si depleted to 7~ppm (limited by measurement noise floor), that is at least 100 nm thick. This is achieved with an ion energy that results in a sputter yield of less than one and an ultra-high ion fluence, as supported by our improved computational model that is based on fitting a large number of experiments. Further, our model predicts the $^{29}$Si and $^{30}$Si depletion in our sample to be less than 1~ppm. In the case of SOI, ion implantation conditions are found to be more stringent than those of bulk natural Si in terms of minimizing threading dislocations upon subsequent solid phase epitaxy annealing. Finally, we do not observe open volume defects in our $^{28}$SOI and $^{28}$Si samples after SPE annealing (620\deg C, 10 minutes). 

\end{abstract}

\maketitle

\section{Introduction}
\label{sec:intro}

Ultra-high fluence $^{28}$Si self-implantation ($\gtrsim 10^{18}$ cm$^{-2}$) can be used as a means to isotopically enrich natural silicon ($^{\rm{nat}}$Si) substrates to unprecedented purity levels. \cite{holmes2021isotopic, acharya2024highly} With the fast growing interest in quantum technology research, $^{28}$Si substrates are important to Si-based quantum information processing (QIP) platforms such as gate-defined quantum dot qubits, \cite{maurand2016cmos, veldhorst2017silicon, gilbert2020single} flip-flop donor qubits, \cite{morello2009architecture, tosi2017silicon, savytskyy2023electrically} and Er-doped Si quantum emitters \cite{yin2013optical, gritsch2022narrow, berkman2025long} due to the zero nuclear spin of $^{28}$Si isotopes and weak spin-orbit coupling in Si. $^{\rm{nat}}$Si consists of 4.7~at.\% of the non-zero nuclear spin isotope $^{29}$Si $(I=1/2)$, leading to decoherence of qubits by spectral diffusion. \cite{witzel2010electron} The other stable isotope, $^{30}$Si, present to 3.1 at.\% has $I=0$ and does not contribute to decoherence via magnetic interactions, however, it can introduce higher order perturbations to the Hamiltonian via strain interactions. \cite{itoh2014isotope} An ideal substrate would therefore consist of isotopically pure monocrystalline $^{28}$Si, the much sought-after `semiconductor spin vacuum'. \cite{steger2012quantum}

The ultra-high fluence $^{28}$Si self-implantation process, which we refer to as the `$^{28}$Si enrichment process', presents itself as a cost-effective and competitive alternative to existing $^{28}$SiF$_4$-based growth techniques \cite{becker2010enrichment, abrosimov2017new, ager2005high, mazzocchi201999, liu202228silicon, itoh2003high, Devyatykh2008} and molecular beam epitaxy (MBE). \cite{klos2024atomistic} $^{28}$SiF$_4$-based techniques rely on toxic and expensive isotopically pure $^{28}$SiF$_4$ gas as its source material. The resultant Si purity is limited by that of the source and can result in purities as low as 50~ppm $^{29}$Si. This material has been used in the Avogadro project to define the kilogram. \cite{becker2012new} A similar level of isotopic purity can also be achieved through solid source MBE (50~ppm $^{29}$Si) which uses $^{28}$Si wafers as its source material. \cite{klos2024atomistic} In contrast, the $^{28}$Si enrichment process employed in the present work uses $^{\rm{nat}}$Si as both its source and starting materials. A $^{\rm{nat}}$Si ion beam is accelerated through a standard magnetic field-based charge-to-mass analyzer and the $^{28}$Si ion beam is mass selected. Through the continuous bombardment of this $^{28}$Si ion beam into a $^{\rm{nat}}$Si substrate, $^{29}$Si and $^{30}$Si are gradually sputtered away, resulting in isotopically purified $^{28}$Si surface layers with a thickness of $\sim100$~nm. \cite{holmes2021isotopic} The ion bombardment results in a transformation to the amorphous silicon phase and a subsequent moderate to high temperature anneal leads to solid phase epitaxial (SPE) regrowth of single crystal silicon with negligible diffusion of the implanted atoms expected. \cite{johnson2015solid, Radek2015, bracht1998silicon} A four to five-fold improvement in the $^{31}$P donor electron spin coherence time as compared to $^{\rm{nat}}$Si has been achieved with a 0.3 at.\% $^{29}$Si (3000 ppm) sample produced by $^{28}$Si enrichment. \cite{holmes2021isotopic} Indeed, the donor coherence time measured in Ref.~\citenum{holland1996implantation} was limited by inter-donor coupling/instantaneous diffusion, thus leading one to expect even longer qubit coherence times in these isotopically enriched $^{28}$Si samples. \cite{bracht1998silicon}

Recently, a $^{29}$Si depletion of 2.3~ppm has been achieved using the $^{28}$Si enrichment process simply by increasing the ion implantation fluence to $10^{19}$ cm$^{-2}$.\cite{acharya2024highly} This will presumably yield an improved semiconductor vacuum and surpass depletion levels attainable with the $^{28}$SiF$_4$-based technique. To achieve such a high purity with the $^{28}$Si enrichment process, a high ion beam flux and fluence are typically required. Previous work in Refs.~\citenum{holmes2021isotopic} and~\citenum{acharya2024highly} employed a flux of \SI{\sim40}{\micro\ampere}/cm$^2$ and \SI{\sim70}{\micro\ampere}/cm$^2$, respectively, and fluences between $10^{18}$ to $10^{19}\rm~cm^{-2}$. However, computational simulations performed in Ref.~\citenum{holmes2021isotopic} were unable to accurately predict the sputtering and depletion characteristics of samples produced from this $^{28}$Si enrichment process. It is also unclear whether the use of high ion fluxes and fluences will negate the benefits of isotopic purity via the formation of microstructural defects such as open volume defects \cite{holmes2021isotopic} as a result of increased damage production rates, localized beam heating, and dynamic annealing during ion implantation. \cite{williams1983production, williams1994ion,
zhu1997microstructure, goldberg1999preferential, holland1996implantation} 

Open volume defects (vacancy clusters, voids) can form in the near-surface region of crystalline Si under non-amorphizing implantation conditions. \cite{brown1998impurity, venezia1998depth, williams2001direct, peeva2000metallic, williams2009voids} This is in part due to the spatial separation between the vacancy and interstitial distributions generated by the ion collisions during implantation. \cite{mazzone1986defect, pellegrino2001separation} An excess of vacancies is expected near the surface up to approximately half the ion projected range ($R_p/2$) while an excess of interstitials is expected deeper in the substrate up to around one to two times the ion projected range ($R_p$ to $2 R_p$). 

For Si implanted under amorphizing conditions, including the $^{28}$Si enrichment process, to our knowledge, there is no direct experimental evidence for the presence of these open volume defects. \cite{holmes2021isotopic, zhu1997microstructure, kudriavtsev2020formation} This may be due to their apparent instability in amorphous Si, \cite{zhu1999instability, zhu2001direct} and resolution limitations of conventional microscopy techniques. The latter may be circumvented through decoration of the open volumes with trace metallic impurities to aid their observation, which may complicate subsequent interpretation. Nevertheless, these open volume defects (if present) will be in close proximity to qubits (usually located within 20 nm of the surface). Thus, it is important to consider the possibility of their formation under the $^{28}$Si enrichment process and their thermal stability under further processing, as explored in this work. 

In this work, we investigate the $^{28}$Si enrichment process using a range of ion beam flux and fluence values. We also consider the impact of ion implantation energy and sample temperature on both the $^{28}$Si enrichment and the crystal quality after SPE annealing. Importantly, the $^{28}$Si enrichment process is also applied to $^{\rm{nat}}$Si-on-insulator ($^{\rm{nat}}$SOI) substrates. SOI substrates are of particular interest for optical interfacing with Er spins \cite{yin2013optical, gritsch2022narrow, berkman2025long} and mechanical resonators \cite{aspelmeyer2014cavity, kumar2022single}, and for gate-defined quantum dot architectures \cite{maurand2016cmos, veldhorst2017silicon} where isotopic purity may be advantageous. We find that, although the $^{28}$Si enrichment process is successfully applied to SOI, a more stringent set of ion implantation conditions must be observed to achieve successful single crystal enriched $^{28}$Si due to flux and temperature effects (collectively referred to as `thermal effects' in this work). This work also leads to an improved computational model of the $^{28}$Si enrichment process, which ultimately enabled us to produce $^{28}$Si samples with $<$7 ppm $^{29}$Si, which is the detection limit of our measurement system. Our model predicts the $^{29}$Si and $^{30}$Si depletion in these samples to be less than 1~ppm.

\section{Experimental and simulation details}
\label{sec:method}

Both $^{\rm nat}$Si substrates (float zone (FZ)-grown, $\rm >10~k\Omega cm$, Topsil) and $^{\rm nat}$SOI substrates with a $(220\pm12.5)$~nm Si device layer (Cz-grown, B-doped) and \SI{3}{\micro\meter} thick buried oxide (Smart-cut) were processed with an ion implanter equipped with a source of negative ions by Cesium sputtering (SNICS) and a 90\deg~mass analyzing magnet (as described in Ref.~\citenum{holmes2021isotopic}). A $^{28}$Si$^-$ ion beam was selected by the magnet, focused to a beam diameter of $\sim3$~mm and raster scanned across a $6\times6$ mm$^2$ square aperture made from intrinsic Si. This prevented any possible forward recoils of undesirable atom species. The beam divergence resulted in an implanted area of $7.5\times7.5$ mm$^2$. The sample chamber vacuum was maintained by a cryopump to between $0.3-3\times10^{-7}$ Torr. The sample stage was tilted by 7\deg~off the ion beam axis and the sample rotated by 15\deg~in the sample stage plane to minimize ion channeling. 

Various implantation parameters were explored including the ion energy and fluence, which were varied between $20-60$ keV and $0.77\times10^{18}-0.77\times10^{19}$ cm$^{-2}$, respectively. The target sample was held at either room temperature (292~K) or liquid nitrogen temperature (77~K). Finally, the ion flux was either maintained above or below \SI{70}{\micro\ampere}/cm$^2$ (obtained by dividing the ion beam current by the ion beam area), and is referred to here as high ($\geq$\SI{70}{\micro\ampere}/cm$^2 = 4.37\times10^{14}$ ions/cm$^2$/s) or low flux, respectively. 

After implantation, samples were cleaned by a degrease (acetone, IPA, DI water), Piranha (H$_2$SO$_4$+H$_2$O$_2$), RCA-2 (HCl+H$_2$O$_2$+H$_2$O), and a dilute HF dip ($< 5\%$ HF). This was proceeded by an SPE anneal using a Modularpro RTP-600xp rapid thermal annealing furnace or a quartz tube furnace at 620\deg C for 10 minutes under an Ar ambient.

Samples were characterized with cross-sectional transmission electron microscopy (TEM, FEI Tecnai F20) and selected area electron diffraction (SAED) on lamellas prepared by Ga$^+$ milling with a focus-ion-beam (ThermoFisher Aquilos 2 cryoFIB). Prior to milling, carbon (C) and platinum (Pt) protective layers were sputter coated and electron and/or ion beam deposited.

The concentration-depth profiles for the three Si isotopes ($^{28}$Si, $^{29}$Si, $^{30}$Si) were obtained by time-of-flight secondary ion mass spectroscopy (ToF-SIMS) with a IONTOF GmbH, TOF.SIMS 5 instrument. A 1~keV Cs$^+$ beam was used for sputtering while a 30~keV Bi$^+$ beam was used for analysis in negative polarity mode. The depth scale was calibrated using the location of the buried Si-SiO$_2$ interface measured from TEM. 

To infer the presence of any open volume defects resulting from the enrichment process, we perform a `gold decoration' experiment as follows. A 0.2~nm thick film of gold (Au) was deposited onto the sample surface with an electron beam evaporator immediately after a diluted HF dip to remove the native oxide. The samples were then subjected to a 620\deg C anneal for 10 minutes in an Ar atmosphere. Samples were then examined with TEM (as detailed above) and Rutherford backscattering spectrometry with channeling analysis (RBS-C). To measure the gold distribution, RBS-C was performed with 1~MeV He$^+$ ions. A Si solid-state detector (18 keV resolution) was placed at an angle of 10\deg~from the incident beam to measure the energy of the backscattered He ions. The portion of the measured RBS spectrum that corresponds to the surface gold distribution was then converted to a depth profile using the RBS analysis software, RUMP. \cite{doolittle1985algorithms} 

The SRIM-2008 simulation package \cite{ziegler2008stopping} was used to calculate the projected ion range, $R_p$, and damage profiles. Surface sputtering and $^{29}$Si depletion analyses were performed with the TRIDYN-2022 simulation package. \cite{moller1988tridyn} Unless specified otherwise, all TRIDYN calculations were performed with the angle of incidence between the ion beam and sample surface normal set to $\alpha=7$\deg and the following simulation control parameters: (1) precision factor $10^{-5}$, (2) relaxation interval 1415, and (3) initial depth interval 1~nm. The substrate was modeled to be a semi-infinite body of Si (single layer) or a two-layered Si/SiO$_2$ stack with natural Si isotopic composition. Thermal and dynamic defect annealing effects arising from varying the substrate temperature and ion flux are not captured by SRIM nor TRIDYN simulations.

\section{Results and discussion}

\subsection{SPE annealing}
\label{sec:spe}

\begin{figure*}[t]
    \centering
    \includegraphics[width=0.9\linewidth]{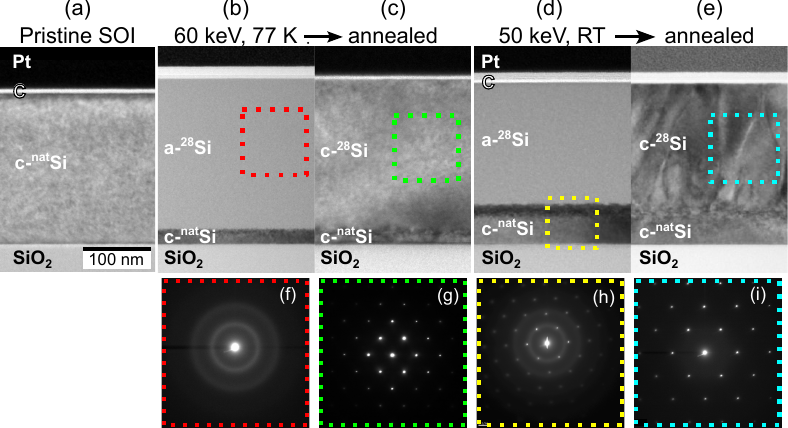}
    \caption{
    TEM images of SOI samples implanted under high flux conditions: (a) Pristine SOI; (b) SOI sample after a 60 keV $0.77\times10^{18}$ cm$^{-2}$ implantation at 77 K and (c) after SPE annealing; (d) SOI sample after a 50 keV $0.77\times10^{18}$ cm$^{-2}$ implantation at RT and (e) after SPE annealing. TEM images are aligned along the Si-SiO$_2$ interface and shared the same scale bar shown in (a). (f-i) SAED images obtained from the dotted regions indicated in the TEM images shown in (b-e).
    }
    \label{fig:tem}
\end{figure*}

TEM images and SAED patterns of selected SOI samples implanted at either 77 K or RT (both under high ion flux conditions) are shown in Fig.~\ref{fig:tem}. A TEM image of pristine SOI is shown for reference [Fig.~\ref{fig:tem}(a)]. Erosion or accumulation caused during the enrichment process are characterized with a change in thickness with reference to this pristine sample, $\Delta z$, where $\Delta z < 0$ indicates erosion, and $\Delta z > 0$ accumulation. All samples in Fig.~\ref{fig:tem} exhibit accumulation with $\Delta z = 16\pm3$ and $\Delta z = 9\pm3$ nm for the samples shown Fig.~\ref{fig:tem}(b) and (d), respectively. The $\Delta z$ of the corresponding samples after SPE annealing [Fig.~\ref{fig:tem}(c) and (e)] are $\Delta z = 15\pm3$ and $\Delta z = 9\pm3$. We note that the density of amorphous Si is 1.8\% less than that of crystalline Si  \cite{custer1994density} and therefore $\Delta z$ in our crystalline Si samples should be 2 to 4 nm less than the corresponding amorphous Si samples. This, however, was not observed in our TEM presumably due to small variations in the Si layer thickness across the SOI substrate as indicated by the supplier. The surface sputter yield, $SY$, quantifies the number of sputtered Si atoms per incident ion and can be calculated assuming negligible changes to the Si atomic density using the equation $SY=1-\frac{\Delta z}{200}\frac{10^{18}}{\Phi}$, where $\Phi$ refers to the implanted $^{28}$Si ion fluence in units of cm$^{-2}$ and $\Delta z$ has units of nm. \footnote{Derived assuming a Si atomic density of $5\times10^{22}$ atoms/cm$^3$ and when the sputter yield is zero, $\Delta z=200$ nm for an ion fluence of $10^{18}$ cm$^{-2}$.} We will discuss the ion flux, fluence and energy dependencies of $\Delta z$ and $SY$ later in Sec.~\ref{sec:delta-z}. 

In Fig.~\ref{fig:tem}(b), it is clear that the ultra-high fluence $^{28}$Si ion implantation (50 keV $0.77\times10^{18}$ cm$^{-2}$, high flux, 77~K) resulted in the formation of a continuous amorphous layer that extends from the surface to almost all the way through the entire Si device layer (labeled `a-$^{28}$Si'). The amorphous nature of this implanted layer is confirmed by SAED measurements (diffuse rings, Fig.~\ref{fig:tem}(f)). Importantly, a 20~nm thick crystalline seed underneath this amorphous layer is retained (labeled `c-$^{\rm{nat}}$Si'). During SPE annealing, this crystal silicon provides a seed for the complete recrystallization of the amorphous layer, as confirmed by TEM [Fig.~\ref{fig:tem}(c)] and SAED measurements (clear bright diffraction spots, Fig.~\ref{fig:tem}(g)). A thin dark band of end-of-range (EOR) interstitial-rich defect complexes is also observed at the same depth location as the original amorphous/crystalline-interface (a/c-interface) [see Figs.~\ref{fig:tem}(b-c)]. Upon SPE annealing at 620\deg C, these interstitials accumulate to form larger and more stable defect complexes/clusters (Ostwald ripening) which can further evolve into loops and rods upon higher temperature annealing at or above 900\deg C. \cite{johnson2015solid} These rods and loop-type defects are visible at the EOR region of our samples under higher magnification TEM [data shown in Supplementary Material]. Other than this region, the recrystallized layer appears to be free from microscopic defects (to within the resolution of our TEM).  

In Fig.~\ref{fig:tem}(d), we show the TEM image of a RT implanted SOI sample (50 keV $0.77\times10^{18}$ cm$^{-2}$, high flux). As compared with the 77 K implanted sample shown previously in Fig.~\ref{fig:tem}(b), the RT implantation also resulted in the formation of a continuous amorphous layer but with a few notable differences. First, the location of the a/c-interface is shallower in the sample shown in Fig.~\ref{fig:tem}(d) as compared to Fig.~\ref{fig:tem}(b) due to a lower ion implantation energy (50 keV and 60 keV, respectively). In general, for sub-100 keV implantation, the location of the a/c-interface is located at a depth of around $2\times R_p$ from the \textit{initial} surface ($\Delta z = 0$), where $R_p$ is the projected ion range calculated from SRIM and is proportional to the ion energy. For example, for $E_{\rm{ion}}=60$ keV, $R_p$ = 87 nm (observed a/c = $(222\pm5)$ nm) while for $E_{\rm{ion}}=50$ keV, $R_p$ = 74 nm (observed a/c = $(179\pm15)$ nm). Second, there is a 20 nm thick dark band of defects located at the broadened a/c-interface in the RT implanted sample (Fig.~\ref{fig:tem}(d)). This is in contrast to the 77 K sample which has a much sharper a/c-interface [Fig.~\ref{fig:tem}(b)]. This dark band of defects may be a result of the accumulation and clustering of a high density of Si interstitials at the a/c-interface during implantation \cite{goldberg1999preferential} and/or dynamic annealing at room temperature which broadens the transition between the amorphous and crystalline phase, as will be discussed later. The TEM image of the same sample after the SPE annealing is shown in Fig.~\ref{fig:tem}(e). Here, threading dislocations were observed to emanate from this dark defect band all the way towards the surface. This was confirmed by TEM where there is evidence for strain. In addition, SAED measurements in the crystalline $^{28}$Si layer show clear bright diffraction spots, as shown in Fig.~\ref{fig:tem}(i).  

The observed threading dislocations are unique to the SOI substrate matrix and are caused by the high ion flux and higher substrate temperature (collectively referred to as `thermal effects'). To elucidate the influence of these thermal effects on the SPE annealing characteristics of our enriched SOI samples, we performed two additional ion implantation experiments: (1) RT implantation on SOI but with a low ion flux and (2) RT and high ion flux implantation on bulk Si. We will describe the results from these two additional experiments here while their associated TEM measurements are shown in the Supplementary Material. In both these cases, the implantation resulted in the formation of an amorphous layer with a sharp a/c-interface. Subsequent SPE annealing resulted in a recrystallized $^{28}$Si layer that is free from microscopic defects and threading dislocations. 

The dependence of the a/c-interface microstructure (sharp or broadened) on the specific implantation conditions (temperature, flux) and substrate type (bulk Si or SOI) can be explained by considering the competition between the generation rate of vacancy-interstitial pairs (a.k.a. Frenkel pairs) and their dynamic annealing. To clarify, dynamic annealing here may refer to the complete annihilation of vacancies and interstitials through a pair-recombination mechanism or their migration to the front/back Si surfaces (referred to as `annihilation'), and/or the formation of larger and less mobile complex defect clusters (referred to as `clustering'). \cite{williams1994ion} Under ion irradiation, Frenkel pairs are generated along the ion track and near the projected ion range in the Si at a rate that is proportional to the ion flux. At 77 K, the mobilities of vacancies and interstitials are reduced as compared to RT. Thus, for the same ion flux, dynamic annealing of point defects at 77 K is suppressed, favoring amorphization. Therefore, for samples implanted under high flux, the a/c-interface is sharp for the 77 K implantation [Fig.~\ref{fig:tem}(b)] due to a reduction in dynamic annealing, while for RT implantation the a/c-interface is broadened [Fig.~\ref{fig:tem}(d)] due to enhanced dynamic annealing (the interstitial clustering rate exceeding its annihilation). In the case where the implantation was performed at RT and under low flux, due to a lower defect generation rate, defect annihilation exceeds its generation and clustering rates and thus the a/c-interface is also sharp [Supplementary Material Fig. S2].

The subsequent SPE annealing characteristics depend crucially on the sharpness of the a/c-interface. The SPE annealing of SOI samples with a sharp or broadened a/c-interface resulted in either high quality single crystal Si free from microscopic defects or threading dislocations, respectively. The latter is most likely caused by an uneven crystalline Si seed plane and/or poor quality of the underlying crystalline Si layer in the presence of a high density of stable interstitial clusters at the a/c-interface as well as strain relaxation effects. \cite{johnson2015solid} The C, N, O impurities might also be expected to influence the nature of the SPE process (e.g., by retarding the SPE rate and resulting in defective growth \cite{johnson2015solid}). However, their presence [see SIMS measurements in Supplementary Material] does not appear to play a role in the formation of these threading dislocations. We also do not expect these threading dislocations to disappear/dissolve completely upon annealing at high temperatures ($>900$\deg C) and/or over a longer duration. Rather, further annealing may result in clustering and Ostwald ripening of the interstitials into more stable defect structures. \cite{johnson2015solid}

The buried Si-SiO$_2$ interface in SOI also plays an important role in the dynamic annealing of interstitial defects generated at or near the a/c-interface. Unlike SOI, when bulk Si samples were implanted at RT under the same high flux conditions, subsequent SPE annealing resulted in the complete recovery of single crystal Si without microscopic defects [Supplementary Material Fig. S2]. This may suggest that the buried oxide in SOI is acting as a physical barrier for mobile interstitials to diffuse and migrate towards the underlying bulk of the substrate, confining the interstitials to a relatively thin Si layer (200 to 300 nm) at the top of the substrate. Further, SOI samples may suffer from a greater degree of localized beam heating \cite{holland1996implantation} due to poorer thermal conduction to the sample stage as compared to bulk Si substrates. This may enhance the formation of stable interstitial clusters during dynamic annealing. 

In summary, the sample temperature and ion flux strongly affect subsequent SPE of enriched $^{28}$SOI samples. This dependence on temperature and flux is unique to the SOI substrate due to the presence of a buried oxide and was not observed in similarly prepared $^{28}$Si samples in bulk Si. For QIP applications, implantation conditions that lead to threading dislocations should be avoided. It remains unclear whether these dislocations will contribute to the surrounding spin bath and lead to further qubit decoherence through spectral diffusion. However, these dislocations are expected to getter undesirable impurities along the threading dislocations within the $^{28}$Si layer which may subsequently form other paramagnetic defect centers. They may also lead to undesirably high diode leakage currents and charge noise, decreased donor activation yields (for example, due to enhanced transient-enhanced diffusion of donors), spatially non-uniform lattice strain fields, and modifications to luminescence spectra. \cite{woo2016insight, johnson2015solid, haunschild2010quality, arguirov2009silicon, stowe2003near}

\subsection{Surface sputtering}
\label{sec:delta-z}

\begin{figure*}[t]
    \centering
    \includegraphics[width=0.8\linewidth]{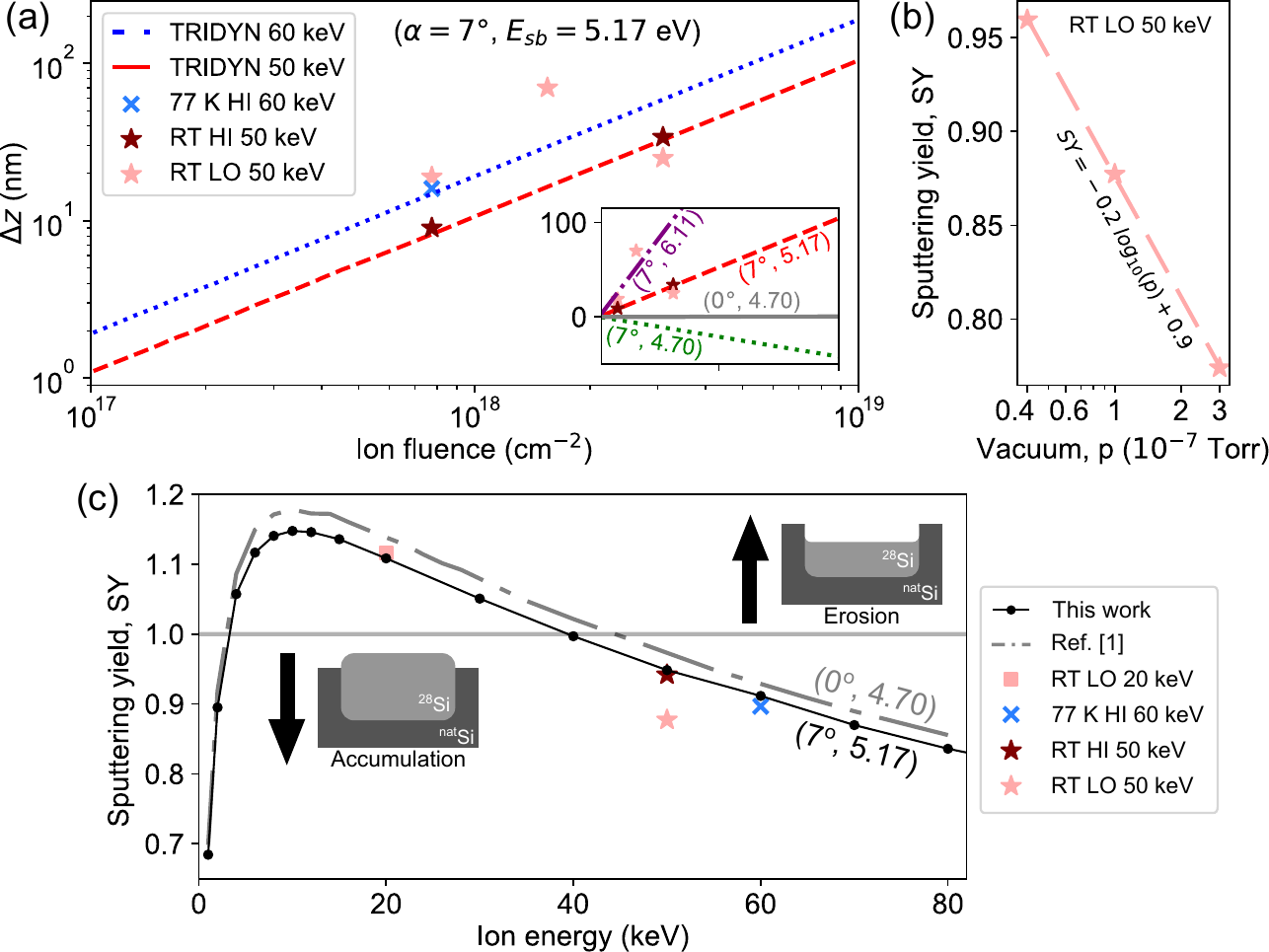}
    \caption{Surface sputtering characteristics after ion implantation: (a) $\Delta z$ as a function of ion fluence calculated for ion energies of 50 keV (red dashed line) and 60 keV (blue dotted line) with $\alpha=7$\deg~and $E_{sb}=5.17$ eV, assuming a two-layered Si/SiO$_2$ stack. Simulations performed for different values of $(\alpha, E_{sb})$ at a fixed ion energy of 50 keV are shown in the inset. Both axes in the inset figure are plotted on a linear scale and the horizontal axis has the same ion fluence range as the main sub-figure. (b) Experimentally measured relationship between the sputtering yield, $SY$, and the sample chamber vacuum level, $p$, for experiments performed under low ion flux conditions at RT. The dashed line is a best fit as indicated by the equation. (c) $SY$ vs. ion energy illustrating both surface accumulation and erosion phenomena. Simulated data are shown as black circles (solid line to guide the eye) and a gray dashed dotted line. The experimental data points (colored) are from samples in this work, implanted up to an ion fluence of $0.77\times10^{18}$ cm$^{-2}$. Experimental error bars are similar in size to the data points ($\sigma_{\text{fluence}} = \pm5\%, \sigma_{\Delta z} = \pm5$ nm). }
    \label{fig:sputter}
\end{figure*}

In Fig.~\ref{fig:sputter}(a), we show TRIDYN simulations of $\Delta z$ as a function of ion fluence fitted to experimental data from this work and plotted on a log scale. The experimentally measured values for $\Delta z$ were obtained through TEM analysis of various SOI samples implanted under different implantation temperatures (RT or 77 K), ion flux (high/HI or low/LO) and energies (50 keV or 60 keV), as described earlier. We note that it was not possible to determine $\Delta z$ for bulk Si substrates from TEM analysis due to the lack of a buried oxide reference but this should be possible through other surface profiling techniques not performed in this work. As for the simulations, the surface binding energy, $E_{sb}$, which is the surface potential that an atom has to overcome to escape the surface, was chosen as the fitting parameter. The simulations shown in Fig.~\ref{fig:sputter}(a) were calculated assuming $E_{sb}=5.17$ eV and a two-layered Si/SiO$_2$ stack. We note that, for 50 keV and 60 keV ions, changing the substrate definition to a single layer of Si increased $SY$ by less than 1\%. This corresponds to a decrease in $\Delta z$, which depends linearly on the ion fluence [see Fig.~\ref{fig:sputter}], by up to 6\% for fluences up to $10^{19}$ cm$^{-2}$. The bulk binding energy, $E_{bulk}$, is set to the recommended value of 0 eV in our simulations, \cite{moller1988tridyn} noting that setting $E_{bulk}=0.1$ eV had a similar effect as increasing $E_{sb}$ by 2 \%. The displacement threshold energy, $E_{dsp}$, had a negligible effect on the simulations performed at 50 keV and 60 keV.

In general, we obtain a reasonable agreement between simulations and our experiments by setting $E_{sb}=5.17$ eV, which is a 10\% increase from its default value of 4.70 eV. As shown in Fig.~\ref{fig:sputter}(a) for 50 keV and 60 keV Si ions, $\Delta z$ is expected to increase linearly with ion fluence. The large discrepancy between the 50 keV simulated curve (red dashed line) and the 50 keV low flux data points (pink stars) in Fig.~\ref{fig:sputter}(a) is due to undesirably high levels of surface contamination from the sample chamber vacuum for these particular implants. Indeed, our experiments indicate that $\Delta z$ becomes increasingly sensitive to the sample chamber vacuum level as the ion flux decreases. As shown in Fig.~\ref{fig:sputter}(b), under low ion flux conditions, we observe a logarithmic relationship between the surface sputtering yield, $SY$, and the sample chamber vacuum level, $p$. A decrease in $SY$ (i.e., increase in $\Delta z$) is observed as the sample chamber background pressure increases, resulting in larger values of $\Delta z$. This is due to the incoming flux of residual gas atoms (contamination) approaching the same order of magnitude as the flux of $^{28}$Si ions. Thus, accurate modeling of $SY$ and $\Delta z$ for a low ion flux and poor vacuum conditions require further adjustments to $E_{sb}$ as illustrated in the inset figure of Fig.~\ref{fig:sputter}(a). In the inset figure, we plot on a linear scale the same relationship between $\Delta z$ and the ion fluence for 50 keV ions but with $E_{sb}$ and $\alpha$ set to different values (labeled as `$(\alpha, E_{sb})$'). 

Vacuum contamination effects are far less significant under high flux conditions and a reasonable agreement between experiment and simulations is achieved by setting $E_{sb}=5.17$ eV as shown in Fig.~\ref{fig:sputter}(a). We believe that the need to \textit{increase} $E_{sb}$ by 10\% from its default value to fit our experimental data is primarily due to low levels of residual gas contamination. In contrast, the implantation temperature and other thermal effects, including local beam heating \cite{holland1996implantation} and poor thermal conduction to the sample stage, are expected to \textit{decrease} $E_{sb}$ by less than $2\%$. This estimate is obtained by assuming that our samples have reached temperatures close to 500--600\deg C during implantation, in which case the thermal energy is $\approx1.5\%$ of $E_{sb}$. Note that it is very unlikely for our samples to have exceeded 600\deg C given the amorphous nature of the implanted layer, as shown previously in Fig.~\ref{fig:tem}, and that the thermal energy near the Si melting point (1400\deg C, $k_B T = 0.15$ eV) is only $3\%$ of $E_{sb}$. Similarly, increased surface damage due to higher ion fluxes can also be expected to \textit{decrease} $E_{sb}$. \cite{sigmund1969theory, almen1961sputtering} Further experiments under high quality vacuum ($<5\times10^{-8}$ Torr) over a wider range of ion fluxes are required to quantify its reducing effect on $E_{sb}$.

In Fig.~\ref{fig:sputter}(c), we show the TRIDYN simulated relationship for $SY$ as a function of the ion energy under a fixed ion fluence of $0.77\times10^{18}$ cm$^{-2}$ and corresponding experimental data from this work. Our simulations (black circles) were performed by setting $\alpha=7$\deg and $E_{sb}=5.17$ eV, and assuming a semi-infinite Si substrate. The black curve is a cubic spline fitted to the simulated data to guide the eye. We note that reducing $E_{sb}$ without changing $\alpha$ shifts the simulated $SY$ curve up without significantly changing its lineshape (data not shown for clarity). The sensitivity of $\Delta z$ and $SY$ to the choice of $\alpha$ and $E_{sb}$ was previously demonstrated in the inset of Fig.~\ref{fig:sputter}(a). Importantly, a sputtering yield of one in our simulated curve occurs at 3.3 keV and 40.0 keV. For comparison, we also show another curve (gray dash-dot line) simulated using the same parameters used in Ref.~\citenum{holmes2021isotopic} ($\alpha=0$\deg, $E_{sb}=4.70$ eV, precision factor 0.1, initial depth interval 0.5 nm, ion fluence $0.77\times10^{18}$ cm$^{-2}$). The one-for-one sputtering mode in this case occurs at 3.0 keV and 44.4 keV ion energies, in good agreement with Ref.~\citenum{holmes2021isotopic}. Clearly, there is a better agreement between our simulated curve (black line) with our experiments. The large discrepancy observed in the 50 keV low flux experiment (pink star) is attributed to vacuum contamination effects discussed previously.

In short, we show that a better agreement between experiments and simulations can be obtained by making small adjustments to $E_{sb}$. In the next section [Sec.\ref{sec:29si}], we show that the $^{29}$Si depletion characteristics from our experiments can also be accurately modeled in TRIDYN by making the same modifications to $E_{sb}$.

\subsection{$^{29}$Si depletion}
\label{sec:29si}

\begin{figure}[t]
    \centering
    \includegraphics[width=1.0\linewidth]{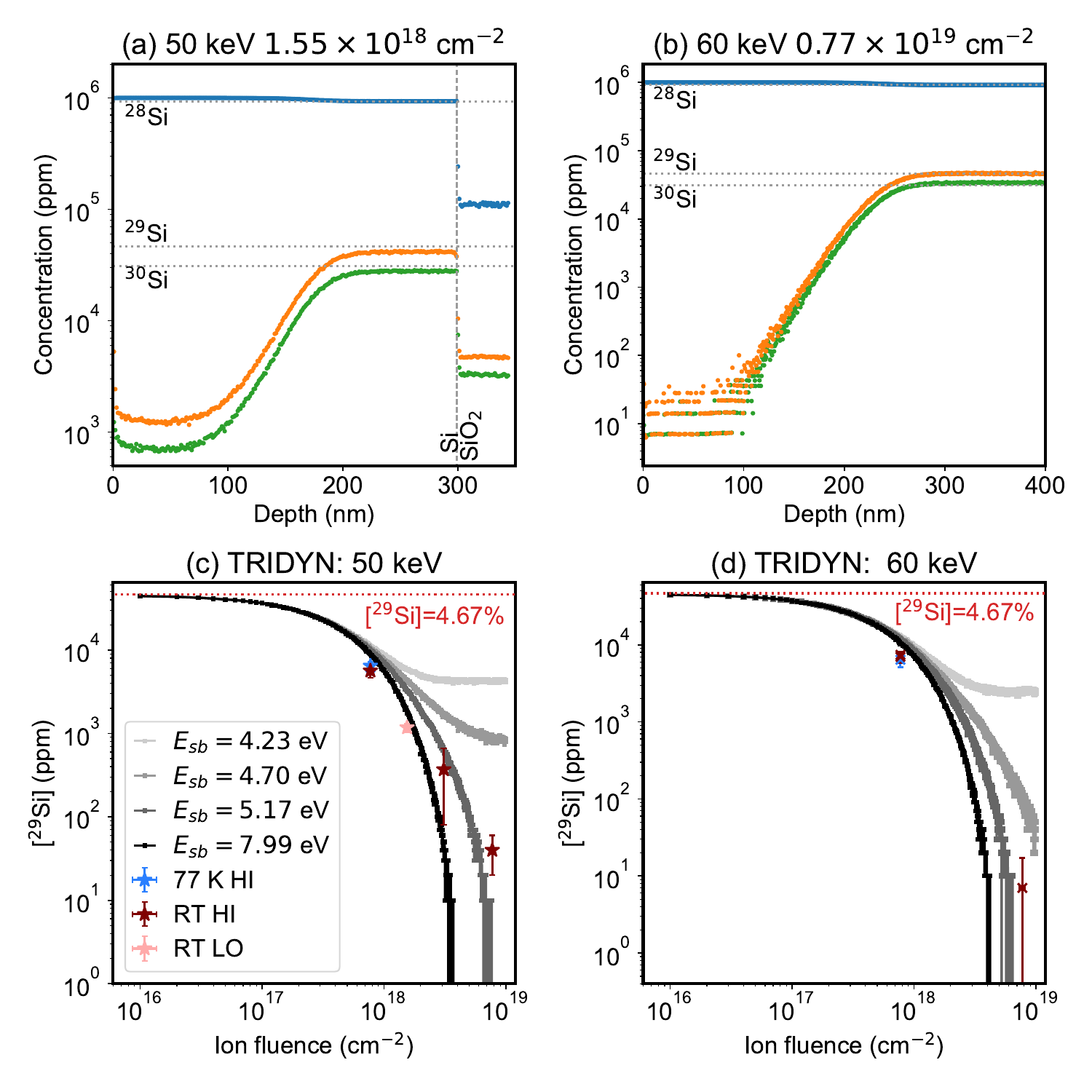}
    \caption{SIMS measured $^{28}$Si, $^{29}$Si, $^{30}$Si isotope profiles from the following samples: (a) 50 keV $1.55\times10^{18}$ cm$^{-2}$, low flux, RT implantation on SOI and (b) 60 keV $0.77\times10^{19}$ cm$^{-2}$, high flux, RT implantation on Si. The vertical dashed line in (a) indicates the buried Si-SiO$_2$ interface. The horizontal dotted lines in (a) and (b) indicate the natural isotopic levels for each Si isotope. TRIDYN calculated [$^{29}$Si] as a function of ion fluence for a fixed ion energy of: (c) 50 keV and, (d) 60 keV, under different values of $E_{sb}$. Experimental data points from this work are also shown (stars and crosses). HI: high flux; LO: low flux. }
    \label{fig:sims}
\end{figure}

The SIMS measured Si isotopic profiles of selected SOI and Si samples are shown in Figs.~\ref{fig:sims}(a) and (b), respectively. First, we note that the sharp concentration spike in the SIMS profiles at the surface is due to a measurement artifact partly related to the native oxide and transient effects. \cite{stevie2005focused} In Fig.~\ref{fig:sims}(a), the abrupt steps in the concentration profiles at a depth location of around 300 nm are due to measurement of the buried oxide interface in the SOI (note $\delta z=70$ nm in this sample) and are indicated by vertical dashed lines and text labels. In all our samples, the region between the surface and a depth of around 100 nm shows significant depletion of both the $^{29}$Si and $^{30}$Si isotopes, with a tail extending to around 200 nm before returning to their natural isotopic levels (or very close to it) indicated by the horizontal dotted lines in Fig.~\ref{fig:sims}(a). Note that SIMS measurements on SOI substrates are more susceptible to surface charging and thus are expected to have larger errors than measurement on Si. We believe that this surface charging effect is responsible for the deviation of the $^{29}$Si and $^{30}$Si concentrations from their natural isotopic level in the deeper regions between 200 nm and 300 nm of the SIMS profiles shown in Fig.~\ref{fig:sims}(a). We define the $^{29}$Si ($^{30}$Si) depletion levels, [$^{29}$Si] ([$^{30}$Si]), as the minimum $^{29}$Si ($^{30}$Si) concentration in the sample; for the sample shown in Fig.~\ref{fig:sims}(a), [$^{29}$Si] = $1200\pm 150$ ppm and [$^{30}$Si] = $700\pm 70$ ppm. On the other hand, the sample shown in Fig.~\ref{fig:sims}(b) is a Si sample implanted under 60 keV $0.77\times10^{19}$ cm$^{-2}$, high flux, RT conditions. Here, both the $^{29}$Si and $^{30}$Si isotopes were depleted below the SIMS instrument noise floor of approximately 7 ppm. We note that 50 nm of the surface of this sample was oxidized during device fabrication experiments prior to SIMS measurement and the Si isotope profiles shown in Fig.~\ref{fig:sims}(b) exclude this oxidized region (no oxide etching performed). Thus, the expected thickness of the $^{28}$Si layer with [$^{29}$Si] and [$^{30}$Si] $<7$ ppm is approximately 150 nm.

In Figs.~\ref{fig:sims}(c) and (d), we plot the TRIDYN curves for [$^{29}$Si] as a function of ion fluence for 50 keV and 60 keV ion energies, simulated under different values of $E_{sb}$. Measurements of [$^{29}$Si] from both SOI and Si samples from our experiments are also shown as scattered data points, where different colors are used to represent the different flux and temperature conditions. A complete data table that summarizes the sample implantation conditions and measured [$^{29}$Si] levels can be found in the Supplementary Material. We first note that the simulations shown here assume a semi-infinite Si medium (single layer). When simulating 50 keV and 60 keV ions, we observed little to no difference in the [$^{29}$Si] vs. ion fluence curves when changing the substrate definition to a two-layered Si/SiO$_2$ stack in TRIDYN. Generally, the experimentally measured Si isotope profiles agree with our TRIDYN simulations within one standard error. Larger deviations between experiment and simulations (especially at the tail of the isotopic profiles) are observed when the ion fluence exceeds $\sim4\times10^{18}$ cm$^{-2}$, presumably due to vacuum-related contamination. A reasonable agreement between data and simulations is obtained when $E_{sb}=5.17$ eV, in good agreement with results from our surface sputtering analyses discussed previously. Importantly, this agreement appears to be independent of the substrate type (Si or SOI). This suggests that local beam heating and thermal conduction to the sample stage play only a minor role (if any) in the $^{29}$Si depletion process. Larger discrepancies between experimental and simulated data are observed in samples that were implanted under low flux conditions (for example, the 50 keV $1.55\times10^{18}$ cm$^{-2}$ data point in Fig.~\ref{fig:sims}(c)) due to vacuum contamination effects discussed previously. 

For a 60 keV $0.77\times10^{19}$ cm$^{-2}$ implantation, TRIDYN predicts a $^{29}$Si depletion of less than 1 ppm, in good agreement with our (noise floor limited) SIMS measurements shown in Fig.~\ref{fig:sims}(b). Additional simulations reveal that $^{29}$Si depletion levels of less than 1 ppm can only be achieved when the ion energy results in surface accumulation. As shown in Fig.~\ref{fig:sputter}(c), surface accumulation in our experiments occurs when the ion energy is less than 3 keV or higher than 40 keV. Under these ion energy conditions, the $^{29}$Si depletion in the sample decreases rapidly as a function of ion fluence and reaches a value of about 1 ppm when the ion fluence is around $0.6\times10^{19}$ to $2.3\times19^{19}$ cm$^{-2}$ [details in Supplementary Material]. Furthermore, by increasing the ion energy from 40 keV to 80 keV, the ion fluence required to achieve $^{29}$Si depletion levels of 1 ppm or less can be lowered from $(2.3\pm0.4)\times10^{19}$ cm$^{-2}$ to $(5.7\pm0.3)\times10^{18}$ cm$^{-2}$. These ion implantation conditions can be easily and realistically achieved as shown by our experiments, demonstrating the competitiveness and attractiveness of our $^{28}$Si enrichment process for producing Si spin vacuum substrates required by QIP. In the next section [Sec.\ref{sec:au}], we show through gold decoration experiments that the isotopically enriched $^{28}$Si samples are free from nanoscopic voids after SPE annealing.

\subsection{Open volume defects}
\label{sec:au}

\begin{figure*}[t]
    \centering
    \includegraphics[width=1.0\linewidth]{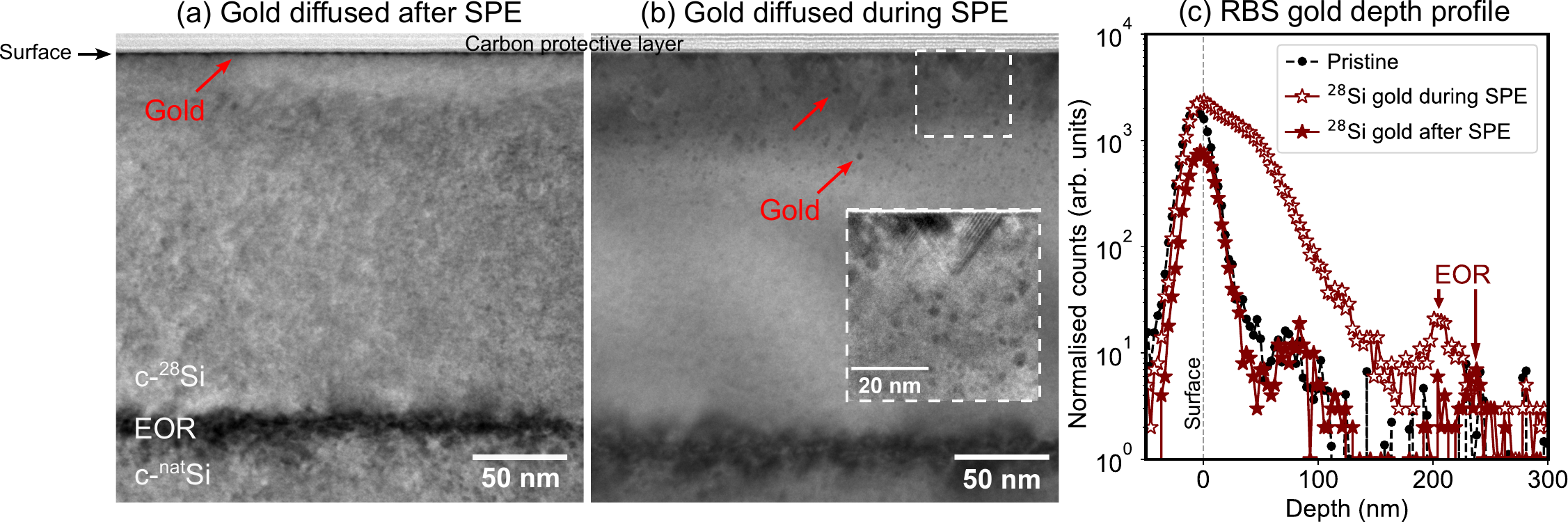}
    \caption{TEM images of $^{28}$Si enriched samples on bulk Si where the gold decoration (diffusion) experiment was performed (a) after and (b) during the SPE anneal stage. The thin carbon protective layer was deposited onto the sample surface during TEM sample preparation. Gold surface segregation and precipitation are indicated by red arrows. A higher magnification TEM image of the region highlighted by the dashed white box is shown as an inset figure. (c) RBS measured gold depth profile for a pristine Si sample after Au diffusion and the $^{28}$Si enriched samples shown in the TEM images (a) and (b). The vertical axis (normalized counts) is proportional to the Au concentration. }
    \label{fig:gold}
\end{figure*}

Figure~\ref{fig:gold}(a) and (b) show TEM results from the $^{28}$Si enriched bulk Si sample (60 keV $0.77\times10^{18}$ cm$^{-2}$, high flux, RT) where the gold in-diffusion anneal was performed either after or during the SPE annealing stage, respectively. Specifically, the sample in Fig.~\ref{fig:gold}(a) was prepared by first performing the SPE anneal (after ion implantation) followed by gold deposition and subsequent annealing to in-diffuse the surface gold. On the other hand, the sample in Fig.~\ref{fig:gold}(b) was prepared by first depositing gold on the sample surface (after ion implantation) followed by SPE annealing. In this case, the gold in-diffusion occurs at the same time as the SPE recrystallization. In both Figs.~\ref{fig:gold}(a) and (b), the dark band of defects located at 200 nm from the surface are the EOR defects, as observed and discussed previously [Fig.~\ref{fig:tem}]. Both of these TEM images were taken under the same magnification settings and the slight difference (7 nm) in the EOR depth location observed between them can be accounted for by TEM scale calibration errors (5\%). 

In the sample where the gold is in-diffused after SPE annealing [Fig.~\ref{fig:gold}(a)], TEM shows a perfectly recrystallized Si layer (supported by SAED measurements, not shown). The gold atoms appear as a thin dark layer segregated at the surface underneath the carbon protective layer. This is confirmed by RBS profiling of the gold that we will discuss later. On the contrary, in the sample where the gold is in-diffused during the SPE annealing stage [Fig.~\ref{fig:gold}(b)], circular regions of darker contrast can be clearly observed up to a depth of 70 nm from the sample surface. These are regions that contain higher concentrations of gold (confirmed by energy dispersive X-ray spectroscopy mapping, data not shown) and are darkened due to the high mass contrast of gold. Two such examples are indicated by the red arrows. We will tentatively refer to them as `gold-rich blobs' and discuss their significance in relation to open volume defects later. The diameter of these gold-rich blobs range from 1 nm to 7 nm, and their size and density distribution achieve a maximum at/near the surface and decrease deeper into the sample. They are no longer visible under TEM at depth regions beyond 70 nm. 

Additionally, we observe stacking-like defects at the surface of the sample in Fig.~\ref{fig:gold}(b). An example of this is highlighted by the white dashed region. A higher magnification TEM image of this region is shown in the inset image. These stacking-like defects were not observed in our other samples that were SPE annealed in the absence of gold at the surface [see, for example, Fig.~\ref{fig:tem}(c) and Fig.~\ref{fig:gold}(a)]. Thus, we attribute these stacking defects to the distortion of the SPE planar growth front due to local modifications to the SPE rate in the presence of gold. \cite{jacobson1986zone} 

RBS-C measured gold depth profiles of the gold diffused samples are shown in Fig.~\ref{fig:gold}(c). A reference gold profile obtained from an un-implanted sample (`Pristine') in-diffused with gold under the same annealing conditions (620\deg C for 10 min) is also shown. The vertical axis (`Normalized counts') is proportional to the gold concentration. In the pristine case, the gold remains predominantly at the surface with a diffusion tail that is 100 times smaller in magnitude that is discernible between the surface and a depth region of 100 nm. We note that the gold surface peak appears slightly broadened at the surface (depth = 0 nm) due to the energy resolution of our RBS detector. 

On the other hand, the gold distribution in the $^{28}$Si samples depends crucially on when the gold in-diffusion was performed. If the gold in-diffusion is performed after the SPE annealing step, we observed no significant difference between the pristine Si and $^{28}$Si gold profiles (solid black circles and red stars, respectively), except a small gold peak that is observed in the $^{28}$Si sample at a depth location greater than 200 nm caused by gold gettering at the EOR defects. The EOR gold gettering peak was observed in both $^{28}$Si samples irrespective of when the gold in-diffusion was performed and is indicated by the arrows and text in Fig.~\ref{fig:gold}(c). The depth locations of the EOR gold gettering peaks observed in RBS ($210\pm20$ nm) are in good agreement with TEM ($220\pm10$ nm). If the gold in-diffusion and SPE annealing are performed simultaneously, we observe in RBS significant accumulation of gold between the surface and 50 nm as compared to the pristine sample, with a concentration tail that decreases and extends deeper into the Si layer. Unsurprisingly, the depth region where the gold is accumulating the most (0 to $50\pm10$ nm) overlaps completely with the TEM depth region where the nanoscopic gold-rich blobs were observed (0 to $70\pm10$ nm). 

We will now discuss the physical origins of these gold-rich blobs and their sub-surface depth distribution. In our pristine Si sample and $^{28}$Si sample where the gold in-diffusion was performed \textit{after} SPE annealing, the in-diffusion of gold occurs entirely in the Si crystalline phase. In crystalline Si, gold diffuses via the kick-out mechanism, where an interstitial gold atom substitutes a Si atom on a lattice site, thereby forming a Si self-interstitial. This diffusion mechanism enables gold to diffuse rapidly in Si, and, combined with a low gold solubility in crystalline Si, results in a U-shaped depth distribution with gold strongly segregating at the front and back surfaces of Si. \cite{gosele1980mechanism} Indeed, this mechanism adequately explains the gold surface segregation observed in both our pristine and $^{28}$Si samples. However, in the $^{28}$Si sample, the perceived gold solubility in the vicinity of the EOR defects (dislocated Si) is higher than in the surrounding (dislocation-free crystalline) Si matrix due to the efficient gold gettering by dislocations at the EOR. \cite{wong1995gettering} 

On the other hand, in the $^{28}$Si sample where the gold was in-diffused \textit{during} the SPE annealing, the initial in-diffusion of gold occurs in the amorphous Si phase. At an annealing temperature of 620\deg C, the diffusion coefficient of gold in amorphous Si and the SPE growth rate are estimated to be 1 nm$^2$/s and 2 nm/s, respectively. \cite{roth1990kinetics, frank1996diffusion} For an initial a/c-interface that is 220 nm deep, the gold in-diffusion front (moving into the substrate) and the SPE growth front (moving towards the surface) will meet at a depth location of 48 nm from the surface after 86 s of annealing. The nature of the gold atoms within the amorphous layer at this stage of the annealing process requires some discussion. Previous works by \citet{elliman1985diffusion} have shown that gold will be uniformly distributed within amorphous Si even at annealing temperatures as low as 450\deg C. Gold precipitation was also observed within the amorphous Si layer following its diffusion, and, curiously, the gold precipitates were localized to only the upper half of the amorphous layer. Indeed, we believe that the gold-rich blobs in Fig.~\ref{fig:gold}(b) are also gold precipitates that were formed in the amorphous Si layer prior to the completion of the SPE. However, rather than their dissolution and zone refinement, \cite{jacobson1986zone} the gold precipitates in our sample remain stable in the crystalline Si phase upon further SPE annealing. We note that zone refinement of gold in amorphous Si is typically expected for soluble gold in its unprecipitated form. It remains unclear whether the gold precipitates in our sample were formed in amorphous Si by decoration/generation of open volumes in the amorphous phase or were formed at the gold-saturated, amorphous-crystalline interface during SPE as a consequence of trapping of some of the segregating gold in the crystalline phase. Further work is needed to elucidate these proposals. 

Finally, it is worth noting that in cases where initially \textit{crystalline} Si is saturated with soluble gold (unprecipitated) through high temperature annealing, subsequent annealing at lower temperatures results in gold precipitation in the crystalline phase as the system evolves to thermal equilibrium (lower gold solubility) at that temperature. \cite{baumann1991precipitation} During this process gold could diffuse via the dissociative mechanism \cite{giese2000microscopic} and precipitate at pre-existing defects or form a series of extrinsic stacking faults. In contrast, gold precipitation in amorphous Si as proposed in this work may be enabled by the presence of high concentrations of gold (and Si dangling bonds) within amorphous Si and thus the average diffusion length of gold in amorphous Si is shorter than the average separation between gold atoms.

\section{Concluding remarks}

Our experiments and simulations demonstrate the ability of the $^{28}$Si enrichment process to precisely control the purity of $^{28}$Si layers formed on bulk Si and SOI substrates. We note that it was necessary to increase the surface binding energy by 10\% to a value of 5.17 eV in TRIDYN to obtain a better agreement with experimentally observed surface sputtering and $^{29}$Si depletion characteristics. Key requirements to produce $^{28}$Si surface layers with residual $^{29}$Si and $^{30}$Si concentrations below 1 ppm and at least 100 nm thick include choosing ion implantation energies that result in surface accumulation ($<3$ keV or $>40$ keV) and ion fluences exceeding $0.6-0.9\times10^{19}$ cm$^{-2}$. We produced one such sample with 60 keV $^{28}$Si ions implanted in bulk Si up to a fluence of $0.77\times10^{19}$ cm$^{-2}$, where the $^{29}$Si and $^{30}$Si isotopes were depleted to less than 7 ppm, a value limited by our SIMS resolution. Further, our experiments emphasize the importance of performing $^{28}$Si enrichment experiments under high vacuum ($<5\times10^{-8}$ Torr) and high ion flux conditions to minimize contamination from background residual gas atoms. High ion fluxes also reduces the overall implantation duration and is desirable for producing samples with larger areas for scaling up. For SOI substrates, high flux ion implantation above $70\ \mu$A/cm$^2$ performed at RT results in significant broadening of the a/c-interface and interstitial clustering which, after subsequent SPE annealing, resulted in the formation of threading dislocations throughout the $^{28}$Si layer. This was not observed in bulk Si substrates and can be circumvented in SOI by performing the ion implantation at 77 K to suppress dynamic annealing. We were unable to detect (with TEM and gold decoration techniques) open volume defects in our $^{28}$Si samples after complete recrystallization by SPE annealing. Future low energy positron annihilation spectroscopy experiments may be better suited at measuring excess vacancies in our samples. 

From a Si device technology and QIP application perspective, our work provides a clear pathway towards realizing large-scale $^{28}$Si and $^{28}$SOI spin vacuum substrates for ultra-long qubit coherence applications. Spin resonance and qubit spectroscopy experiments are currently underway to uncover the coherent lifetimes of donor qubits embedded in our $^{28}$Si and $^{28}$SOI substrates.

\section{Acknowledgments}

This research was supported by the Australian Government through the Australian Research Council Centre of Excellence for Quantum Computation and Communication Technology (CE170100012) and Discovery Projects funding scheme (DP220103467). We acknowledge the use of the NCRIS Heavy Ion Accelerator platform (HIA) for access and support to the low energy implanter equipment at the Australian National University, the Bio21 institute for the use of FIB and TEM facilities at the University of Melbourne, and the surface analysis laboratory (SSEAU, MWAC) for access and support to the SIMS equipment at UNSW. TRIDYN calculations were performed with multi-thread parallelization using the Spartan High Performance Computing (HPC) system supported by the University of Melbourne's Research Computing Services and the Petascale Campus Initiative. SQL and JCM thank Wolfhard Moller, Benoit Voisin,  Gabriele de Boo, Alexey Lyasota, and Sven Rogge for useful discussions. We are grateful to Juha Mohonen for providing the SOI substrates.


\begin{thebibliography}{63}%
\makeatletter
\providecommand \@ifxundefined [1]{%
 \@ifx{#1\undefined}
}%
\providecommand \@ifnum [1]{%
 \ifnum #1\expandafter \@firstoftwo
 \else \expandafter \@secondoftwo
 \fi
}%
\providecommand \@ifx [1]{%
 \ifx #1\expandafter \@firstoftwo
 \else \expandafter \@secondoftwo
 \fi
}%
\providecommand \natexlab [1]{#1}%
\providecommand \enquote  [1]{``#1''}%
\providecommand \bibnamefont  [1]{#1}%
\providecommand \bibfnamefont [1]{#1}%
\providecommand \citenamefont [1]{#1}%
\providecommand \href@noop [0]{\@secondoftwo}%
\providecommand \href [0]{\begingroup \@sanitize@url \@href}%
\providecommand \@href[1]{\@@startlink{#1}\@@href}%
\providecommand \@@href[1]{\endgroup#1\@@endlink}%
\providecommand \@sanitize@url [0]{\catcode `\\12\catcode `\$12\catcode `\&12\catcode `\#12\catcode `\^12\catcode `\_12\catcode `\%12\relax}%
\providecommand \@@startlink[1]{}%
\providecommand \@@endlink[0]{}%
\providecommand \url  [0]{\begingroup\@sanitize@url \@url }%
\providecommand \@url [1]{\endgroup\@href {#1}{\urlprefix }}%
\providecommand \urlprefix  [0]{URL }%
\providecommand \Eprint [0]{\href }%
\providecommand \doibase [0]{https://doi.org/}%
\providecommand \selectlanguage [0]{\@gobble}%
\providecommand \bibinfo  [0]{\@secondoftwo}%
\providecommand \bibfield  [0]{\@secondoftwo}%
\providecommand \translation [1]{[#1]}%
\providecommand \BibitemOpen [0]{}%
\providecommand \bibitemStop [0]{}%
\providecommand \bibitemNoStop [0]{.\EOS\space}%
\providecommand \EOS [0]{\spacefactor3000\relax}%
\providecommand \BibitemShut  [1]{\csname bibitem#1\endcsname}%
\let\auto@bib@innerbib\@empty
%</preamble>
\bibitem [{\citenamefont {Holmes}\ \emph {et~al.}(2021)\citenamefont {Holmes}, \citenamefont {Johnson}, \citenamefont {Chua}, \citenamefont {Voisin}, \citenamefont {Kocsis}, \citenamefont {Rubanov}, \citenamefont {Robson}, \citenamefont {McCallum}, \citenamefont {McCamey}, \citenamefont {Rogge} \emph {et~al.}}]{holmes2021isotopic}%
  \BibitemOpen
  \bibfield  {author} {\bibinfo {author} {\bibfnamefont {D.}~\bibnamefont {Holmes}}, \bibinfo {author} {\bibfnamefont {B.~C.}\ \bibnamefont {Johnson}}, \bibinfo {author} {\bibfnamefont {C.}~\bibnamefont {Chua}}, \bibinfo {author} {\bibfnamefont {B.}~\bibnamefont {Voisin}}, \bibinfo {author} {\bibfnamefont {S.}~\bibnamefont {Kocsis}}, \bibinfo {author} {\bibfnamefont {S.}~\bibnamefont {Rubanov}}, \bibinfo {author} {\bibfnamefont {S.}~\bibnamefont {Robson}}, \bibinfo {author} {\bibfnamefont {J.~C.}\ \bibnamefont {McCallum}}, \bibinfo {author} {\bibfnamefont {D.~R.}\ \bibnamefont {McCamey}}, \bibinfo {author} {\bibfnamefont {S.}~\bibnamefont {Rogge}}, \emph {et~al.},\ }\bibfield  {title} {\bibinfo {title} {{Isotopic enrichment of silicon by high fluence 28 Si- ion implantation}},\ }\href@noop {} {\bibfield  {journal} {\bibinfo  {journal} {{Physical Review Materials}}\ }\textbf {\bibinfo {volume} {5}},\ \bibinfo {pages} {014601} (\bibinfo {year} {2021})}\BibitemShut {NoStop}%
\bibitem [{\citenamefont {Acharya}\ \emph {et~al.}(2024)\citenamefont {Acharya}, \citenamefont {Coke}, \citenamefont {Adshead}, \citenamefont {Li}, \citenamefont {Achinuq}, \citenamefont {Cai}, \citenamefont {Gholizadeh}, \citenamefont {Jacobs}, \citenamefont {Boland}, \citenamefont {Haigh}, \citenamefont {Boland}, \citenamefont {Haigh}, \citenamefont {Moore}, \citenamefont {Jamieson},\ and\ \citenamefont {Curry}}]{acharya2024highly}%
  \BibitemOpen
  \bibfield  {author} {\bibinfo {author} {\bibfnamefont {R.}~\bibnamefont {Acharya}}, \bibinfo {author} {\bibfnamefont {M.}~\bibnamefont {Coke}}, \bibinfo {author} {\bibfnamefont {M.}~\bibnamefont {Adshead}}, \bibinfo {author} {\bibfnamefont {K.}~\bibnamefont {Li}}, \bibinfo {author} {\bibfnamefont {B.}~\bibnamefont {Achinuq}}, \bibinfo {author} {\bibfnamefont {R.}~\bibnamefont {Cai}}, \bibinfo {author} {\bibfnamefont {A.~B.}\ \bibnamefont {Gholizadeh}}, \bibinfo {author} {\bibfnamefont {J.}~\bibnamefont {Jacobs}}, \bibinfo {author} {\bibfnamefont {J.~L.}\ \bibnamefont {Boland}}, \bibinfo {author} {\bibfnamefont {S.~J.}\ \bibnamefont {Haigh}}, \bibinfo {author} {\bibfnamefont {J.~L.}\ \bibnamefont {Boland}}, \bibinfo {author} {\bibfnamefont {S.~J.}\ \bibnamefont {Haigh}}, \bibinfo {author} {\bibfnamefont {K.~L.}\ \bibnamefont {Moore}}, \bibinfo {author} {\bibfnamefont {D.~N.}\ \bibnamefont {Jamieson}},\ and\ \bibinfo {author} {\bibfnamefont {R.~J.}\ \bibnamefont {Curry}},\ }\bibfield  {title} {\bibinfo
  {title} {{Highly 28Si enriched silicon by localised focused ion beam implantation}},\ }\href@noop {} {\bibfield  {journal} {\bibinfo  {journal} {{Communications Materials}}\ }\textbf {\bibinfo {volume} {5}},\ \bibinfo {pages} {57} (\bibinfo {year} {2024})}\BibitemShut {NoStop}%
\bibitem [{\citenamefont {Maurand}\ \emph {et~al.}(2016)\citenamefont {Maurand}, \citenamefont {Jehl}, \citenamefont {Kotekar-Patil}, \citenamefont {Corna}, \citenamefont {Bohuslavskyi}, \citenamefont {Lavi{\'e}ville}, \citenamefont {Hutin}, \citenamefont {Barraud}, \citenamefont {Vinet}, \citenamefont {Sanquer} \emph {et~al.}}]{maurand2016cmos}%
  \BibitemOpen
  \bibfield  {author} {\bibinfo {author} {\bibfnamefont {R.}~\bibnamefont {Maurand}}, \bibinfo {author} {\bibfnamefont {X.}~\bibnamefont {Jehl}}, \bibinfo {author} {\bibfnamefont {D.}~\bibnamefont {Kotekar-Patil}}, \bibinfo {author} {\bibfnamefont {A.}~\bibnamefont {Corna}}, \bibinfo {author} {\bibfnamefont {H.}~\bibnamefont {Bohuslavskyi}}, \bibinfo {author} {\bibfnamefont {R.}~\bibnamefont {Lavi{\'e}ville}}, \bibinfo {author} {\bibfnamefont {L.}~\bibnamefont {Hutin}}, \bibinfo {author} {\bibfnamefont {S.}~\bibnamefont {Barraud}}, \bibinfo {author} {\bibfnamefont {M.}~\bibnamefont {Vinet}}, \bibinfo {author} {\bibfnamefont {M.}~\bibnamefont {Sanquer}}, \emph {et~al.},\ }\bibfield  {title} {\bibinfo {title} {{A CMOS silicon spin qubit}},\ }\href@noop {} {\bibfield  {journal} {\bibinfo  {journal} {{Nature communications}}\ }\textbf {\bibinfo {volume} {7}},\ \bibinfo {pages} {13575} (\bibinfo {year} {2016})}\BibitemShut {NoStop}%
\bibitem [{\citenamefont {Veldhorst}\ \emph {et~al.}(2017)\citenamefont {Veldhorst}, \citenamefont {Eenink}, \citenamefont {Yang},\ and\ \citenamefont {Dzurak}}]{veldhorst2017silicon}%
  \BibitemOpen
  \bibfield  {author} {\bibinfo {author} {\bibfnamefont {M.}~\bibnamefont {Veldhorst}}, \bibinfo {author} {\bibfnamefont {H.~G.~J.}\ \bibnamefont {Eenink}}, \bibinfo {author} {\bibfnamefont {C.-H.}\ \bibnamefont {Yang}},\ and\ \bibinfo {author} {\bibfnamefont {A.~S.}\ \bibnamefont {Dzurak}},\ }\bibfield  {title} {\bibinfo {title} {{Silicon CMOS architecture for a spin-based quantum computer}},\ }\href@noop {} {\bibfield  {journal} {\bibinfo  {journal} {{Nature communications}}\ }\textbf {\bibinfo {volume} {8}},\ \bibinfo {pages} {1766} (\bibinfo {year} {2017})}\BibitemShut {NoStop}%
\bibitem [{\citenamefont {Gilbert}\ \emph {et~al.}(2020)\citenamefont {Gilbert}, \citenamefont {Saraiva}, \citenamefont {Lim}, \citenamefont {Yang}, \citenamefont {Laucht}, \citenamefont {Bertrand}, \citenamefont {Rambal}, \citenamefont {Hutin}, \citenamefont {Escott}, \citenamefont {Vinet},\ and\ \citenamefont {Dzurak}}]{gilbert2020single}%
  \BibitemOpen
  \bibfield  {author} {\bibinfo {author} {\bibfnamefont {W.}~\bibnamefont {Gilbert}}, \bibinfo {author} {\bibfnamefont {A.}~\bibnamefont {Saraiva}}, \bibinfo {author} {\bibfnamefont {W.~H.}\ \bibnamefont {Lim}}, \bibinfo {author} {\bibfnamefont {C.~H.}\ \bibnamefont {Yang}}, \bibinfo {author} {\bibfnamefont {A.}~\bibnamefont {Laucht}}, \bibinfo {author} {\bibfnamefont {B.}~\bibnamefont {Bertrand}}, \bibinfo {author} {\bibfnamefont {N.}~\bibnamefont {Rambal}}, \bibinfo {author} {\bibfnamefont {L.}~\bibnamefont {Hutin}}, \bibinfo {author} {\bibfnamefont {C.~C.}\ \bibnamefont {Escott}}, \bibinfo {author} {\bibfnamefont {M.}~\bibnamefont {Vinet}},\ and\ \bibinfo {author} {\bibfnamefont {A.~S.}\ \bibnamefont {Dzurak}},\ }\bibfield  {title} {\bibinfo {title} {{Single-electron operation of a silicon-CMOS 2$\times$ 2 quantum dot array with integrated charge sensing}},\ }\href@noop {} {\bibfield  {journal} {\bibinfo  {journal} {{Nano Letters}}\ }\textbf {\bibinfo {volume} {20}},\ \bibinfo {pages} {7882} (\bibinfo
  {year} {2020})}\BibitemShut {NoStop}%
\bibitem [{\citenamefont {Morello}\ \emph {et~al.}(2009)\citenamefont {Morello}, \citenamefont {Escott}, \citenamefont {Huebl}, \citenamefont {Willems~van Beveren}, \citenamefont {Hollenberg}, \citenamefont {Jamieson}, \citenamefont {Dzurak},\ and\ \citenamefont {Clark}}]{morello2009architecture}%
  \BibitemOpen
  \bibfield  {author} {\bibinfo {author} {\bibfnamefont {A.}~\bibnamefont {Morello}}, \bibinfo {author} {\bibfnamefont {C.~C.}\ \bibnamefont {Escott}}, \bibinfo {author} {\bibfnamefont {H.}~\bibnamefont {Huebl}}, \bibinfo {author} {\bibfnamefont {L.~H.}\ \bibnamefont {Willems~van Beveren}}, \bibinfo {author} {\bibfnamefont {L.~C.~L.}\ \bibnamefont {Hollenberg}}, \bibinfo {author} {\bibfnamefont {D.~N.}\ \bibnamefont {Jamieson}}, \bibinfo {author} {\bibfnamefont {A.~S.}\ \bibnamefont {Dzurak}},\ and\ \bibinfo {author} {\bibfnamefont {R.~G.}\ \bibnamefont {Clark}},\ }\bibfield  {title} {\bibinfo {title} {{Architecture for high-sensitivity single-shot readout and control of the electron spin of individual donors in silicon}},\ }\href@noop {} {\bibfield  {journal} {\bibinfo  {journal} {{Physical Review B—Condensed Matter and Materials Physics}}\ }\textbf {\bibinfo {volume} {80}},\ \bibinfo {pages} {081307} (\bibinfo {year} {2009})}\BibitemShut {NoStop}%
\bibitem [{\citenamefont {Tosi}\ \emph {et~al.}(2017)\citenamefont {Tosi}, \citenamefont {Mohiyaddin}, \citenamefont {Schmitt}, \citenamefont {Tenberg}, \citenamefont {Rahman}, \citenamefont {Klimeck},\ and\ \citenamefont {Morello}}]{tosi2017silicon}%
  \BibitemOpen
  \bibfield  {author} {\bibinfo {author} {\bibfnamefont {G.}~\bibnamefont {Tosi}}, \bibinfo {author} {\bibfnamefont {F.~A.}\ \bibnamefont {Mohiyaddin}}, \bibinfo {author} {\bibfnamefont {V.}~\bibnamefont {Schmitt}}, \bibinfo {author} {\bibfnamefont {S.}~\bibnamefont {Tenberg}}, \bibinfo {author} {\bibfnamefont {R.}~\bibnamefont {Rahman}}, \bibinfo {author} {\bibfnamefont {G.}~\bibnamefont {Klimeck}},\ and\ \bibinfo {author} {\bibfnamefont {A.}~\bibnamefont {Morello}},\ }\bibfield  {title} {\bibinfo {title} {{Silicon quantum processor with robust long-distance qubit couplings}},\ }\href@noop {} {\bibfield  {journal} {\bibinfo  {journal} {{Nature communications}}\ }\textbf {\bibinfo {volume} {8}},\ \bibinfo {pages} {450} (\bibinfo {year} {2017})}\BibitemShut {NoStop}%
\bibitem [{\citenamefont {Savytskyy}\ \emph {et~al.}(2023)\citenamefont {Savytskyy}, \citenamefont {Botzem}, \citenamefont {Fernandez~de Fuentes}, \citenamefont {Joecker}, \citenamefont {Pla}, \citenamefont {Hudson}, \citenamefont {Itoh}, \citenamefont {Jakob}, \citenamefont {Johnson}, \citenamefont {Jamieson}, \citenamefont {Dzurak},\ and\ \citenamefont {Morello}}]{savytskyy2023electrically}%
  \BibitemOpen
  \bibfield  {author} {\bibinfo {author} {\bibfnamefont {R.}~\bibnamefont {Savytskyy}}, \bibinfo {author} {\bibfnamefont {T.}~\bibnamefont {Botzem}}, \bibinfo {author} {\bibfnamefont {I.}~\bibnamefont {Fernandez~de Fuentes}}, \bibinfo {author} {\bibfnamefont {B.}~\bibnamefont {Joecker}}, \bibinfo {author} {\bibfnamefont {J.~J.}\ \bibnamefont {Pla}}, \bibinfo {author} {\bibfnamefont {F.~E.}\ \bibnamefont {Hudson}}, \bibinfo {author} {\bibfnamefont {K.~M.}\ \bibnamefont {Itoh}}, \bibinfo {author} {\bibfnamefont {A.~M.}\ \bibnamefont {Jakob}}, \bibinfo {author} {\bibfnamefont {B.~C.}\ \bibnamefont {Johnson}}, \bibinfo {author} {\bibfnamefont {D.~N.}\ \bibnamefont {Jamieson}}, \bibinfo {author} {\bibfnamefont {A.~S.}\ \bibnamefont {Dzurak}},\ and\ \bibinfo {author} {\bibfnamefont {A.}~\bibnamefont {Morello}},\ }\bibfield  {title} {\bibinfo {title} {{An electrically driven single-atom “flip-flop” qubit}},\ }\href@noop {} {\bibfield  {journal} {\bibinfo  {journal} {{Science advances}}\ }\textbf {\bibinfo
  {volume} {9}},\ \bibinfo {pages} {eadd9408} (\bibinfo {year} {2023})}\BibitemShut {NoStop}%
\bibitem [{\citenamefont {Yin}\ \emph {et~al.}(2013)\citenamefont {Yin}, \citenamefont {Rancic}, \citenamefont {de~Boo}, \citenamefont {Stavrias}, \citenamefont {McCallum}, \citenamefont {Sellars},\ and\ \citenamefont {Rogge}}]{yin2013optical}%
  \BibitemOpen
  \bibfield  {author} {\bibinfo {author} {\bibfnamefont {C.}~\bibnamefont {Yin}}, \bibinfo {author} {\bibfnamefont {M.}~\bibnamefont {Rancic}}, \bibinfo {author} {\bibfnamefont {G.~G.}\ \bibnamefont {de~Boo}}, \bibinfo {author} {\bibfnamefont {N.}~\bibnamefont {Stavrias}}, \bibinfo {author} {\bibfnamefont {J.~C.}\ \bibnamefont {McCallum}}, \bibinfo {author} {\bibfnamefont {M.~J.}\ \bibnamefont {Sellars}},\ and\ \bibinfo {author} {\bibfnamefont {S.}~\bibnamefont {Rogge}},\ }\bibfield  {title} {\bibinfo {title} {{Optical addressing of an individual erbium ion in silicon}},\ }\href@noop {} {\bibfield  {journal} {\bibinfo  {journal} {{Nature}}\ }\textbf {\bibinfo {volume} {497}},\ \bibinfo {pages} {91} (\bibinfo {year} {2013})}\BibitemShut {NoStop}%
\bibitem [{\citenamefont {Gritsch}\ \emph {et~al.}(2022)\citenamefont {Gritsch}, \citenamefont {Weiss}, \citenamefont {Fr{\"u}h}, \citenamefont {Rinner},\ and\ \citenamefont {Reiserer}}]{gritsch2022narrow}%
  \BibitemOpen
  \bibfield  {author} {\bibinfo {author} {\bibfnamefont {A.}~\bibnamefont {Gritsch}}, \bibinfo {author} {\bibfnamefont {L.}~\bibnamefont {Weiss}}, \bibinfo {author} {\bibfnamefont {J.}~\bibnamefont {Fr{\"u}h}}, \bibinfo {author} {\bibfnamefont {S.}~\bibnamefont {Rinner}},\ and\ \bibinfo {author} {\bibfnamefont {A.}~\bibnamefont {Reiserer}},\ }\bibfield  {title} {\bibinfo {title} {{Narrow optical transitions in erbium-implanted silicon waveguides}},\ }\href@noop {} {\bibfield  {journal} {\bibinfo  {journal} {{Physical Review X}}\ }\textbf {\bibinfo {volume} {12}},\ \bibinfo {pages} {041009} (\bibinfo {year} {2022})}\BibitemShut {NoStop}%
\bibitem [{\citenamefont {Berkman}\ \emph {et~al.}(2025)\citenamefont {Berkman}, \citenamefont {Lyasota}, \citenamefont {de~Boo}, \citenamefont {Bartholomew}, \citenamefont {Lim}, \citenamefont {Johnson}, \citenamefont {McCallum}, \citenamefont {Xu}, \citenamefont {Xie}, \citenamefont {Abrosimov}, \citenamefont {Pohlm}, \citenamefont {Ahlefeldt}, \citenamefont {Sellars}, \citenamefont {Yin},\ and\ \citenamefont {Rogge}}]{berkman2025long}%
  \BibitemOpen
  \bibfield  {author} {\bibinfo {author} {\bibfnamefont {I.~R.}\ \bibnamefont {Berkman}}, \bibinfo {author} {\bibfnamefont {A.}~\bibnamefont {Lyasota}}, \bibinfo {author} {\bibfnamefont {G.~G.}\ \bibnamefont {de~Boo}}, \bibinfo {author} {\bibfnamefont {J.~G.}\ \bibnamefont {Bartholomew}}, \bibinfo {author} {\bibfnamefont {S.~Q.}\ \bibnamefont {Lim}}, \bibinfo {author} {\bibfnamefont {B.~C.}\ \bibnamefont {Johnson}}, \bibinfo {author} {\bibfnamefont {J.~C.}\ \bibnamefont {McCallum}}, \bibinfo {author} {\bibfnamefont {B.-B.}\ \bibnamefont {Xu}}, \bibinfo {author} {\bibfnamefont {S.}~\bibnamefont {Xie}}, \bibinfo {author} {\bibfnamefont {N.~V.}\ \bibnamefont {Abrosimov}}, \bibinfo {author} {\bibfnamefont {H.-J.}\ \bibnamefont {Pohlm}}, \bibinfo {author} {\bibfnamefont {R.~L.}\ \bibnamefont {Ahlefeldt}}, \bibinfo {author} {\bibfnamefont {M.~J.}\ \bibnamefont {Sellars}}, \bibinfo {author} {\bibfnamefont {C.}~\bibnamefont {Yin}},\ and\ \bibinfo {author} {\bibfnamefont {S.}~\bibnamefont {Rogge}},\ }\bibfield
  {title} {\bibinfo {title} {Long optical and electron spin coherence times for erbium ions in silicon},\ }\href@noop {} {\bibfield  {journal} {\bibinfo  {journal} {{npj Quantum Information}}\ }\textbf {\bibinfo {volume} {11}},\ \bibinfo {pages} {66} (\bibinfo {year} {2025})}\BibitemShut {NoStop}%
\bibitem [{\citenamefont {Witzel}\ \emph {et~al.}(2010)\citenamefont {Witzel}, \citenamefont {Carroll}, \citenamefont {Morello}, \citenamefont {Cywi{\'n}ski},\ and\ \citenamefont {Das~Sarma}}]{witzel2010electron}%
  \BibitemOpen
  \bibfield  {author} {\bibinfo {author} {\bibfnamefont {W.~M.}\ \bibnamefont {Witzel}}, \bibinfo {author} {\bibfnamefont {M.~S.}\ \bibnamefont {Carroll}}, \bibinfo {author} {\bibfnamefont {A.}~\bibnamefont {Morello}}, \bibinfo {author} {\bibfnamefont {{\L}.}~\bibnamefont {Cywi{\'n}ski}},\ and\ \bibinfo {author} {\bibfnamefont {S.}~\bibnamefont {Das~Sarma}},\ }\bibfield  {title} {\bibinfo {title} {{Electron spin decoherence in isotope-enriched silicon}},\ }\href@noop {} {\bibfield  {journal} {\bibinfo  {journal} {{Physical Review Letters}}\ }\textbf {\bibinfo {volume} {105}},\ \bibinfo {pages} {187602} (\bibinfo {year} {2010})}\BibitemShut {NoStop}%
\bibitem [{\citenamefont {Itoh}\ and\ \citenamefont {Watanabe}(2014)}]{itoh2014isotope}%
  \BibitemOpen
  \bibfield  {author} {\bibinfo {author} {\bibfnamefont {K.~M.}\ \bibnamefont {Itoh}}\ and\ \bibinfo {author} {\bibfnamefont {H.}~\bibnamefont {Watanabe}},\ }\bibfield  {title} {\bibinfo {title} {{Isotope engineering of silicon and diamond for quantum computing and sensing applications}},\ }\href@noop {} {\bibfield  {journal} {\bibinfo  {journal} {{MRS communications}}\ }\textbf {\bibinfo {volume} {4}},\ \bibinfo {pages} {143} (\bibinfo {year} {2014})}\BibitemShut {NoStop}%
\bibitem [{\citenamefont {Steger}\ \emph {et~al.}(2012)\citenamefont {Steger}, \citenamefont {Saeedi}, \citenamefont {Thewalt}, \citenamefont {Morton}, \citenamefont {Riemann}, \citenamefont {Abrosimov}, \citenamefont {Becker},\ and\ \citenamefont {Pohl}}]{steger2012quantum}%
  \BibitemOpen
  \bibfield  {author} {\bibinfo {author} {\bibfnamefont {M.}~\bibnamefont {Steger}}, \bibinfo {author} {\bibfnamefont {K.}~\bibnamefont {Saeedi}}, \bibinfo {author} {\bibfnamefont {M.~L.~W.}\ \bibnamefont {Thewalt}}, \bibinfo {author} {\bibfnamefont {J.~J.~L.}\ \bibnamefont {Morton}}, \bibinfo {author} {\bibfnamefont {H.}~\bibnamefont {Riemann}}, \bibinfo {author} {\bibfnamefont {N.~V.}\ \bibnamefont {Abrosimov}}, \bibinfo {author} {\bibfnamefont {P.}~\bibnamefont {Becker}},\ and\ \bibinfo {author} {\bibfnamefont {H.-J.}\ \bibnamefont {Pohl}},\ }\bibfield  {title} {\bibinfo {title} {Quantum information storage for over 180 s using donor spins in a 28si “semiconductor vacuum”},\ }\href@noop {} {\bibfield  {journal} {\bibinfo  {journal} {Science}\ }\textbf {\bibinfo {volume} {336}},\ \bibinfo {pages} {1280} (\bibinfo {year} {2012})}\BibitemShut {NoStop}%
\bibitem [{\citenamefont {Becker}\ \emph {et~al.}(2010)\citenamefont {Becker}, \citenamefont {Pohl}, \citenamefont {Riemann},\ and\ \citenamefont {Abrosimov}}]{becker2010enrichment}%
  \BibitemOpen
  \bibfield  {author} {\bibinfo {author} {\bibfnamefont {P.}~\bibnamefont {Becker}}, \bibinfo {author} {\bibfnamefont {H.-J.}\ \bibnamefont {Pohl}}, \bibinfo {author} {\bibfnamefont {H.}~\bibnamefont {Riemann}},\ and\ \bibinfo {author} {\bibfnamefont {N.}~\bibnamefont {Abrosimov}},\ }\bibfield  {title} {\bibinfo {title} {{Enrichment of silicon for a better kilogram}},\ }\href@noop {} {\bibfield  {journal} {\bibinfo  {journal} {{physica status solidi (a)}}\ }\textbf {\bibinfo {volume} {207}},\ \bibinfo {pages} {49} (\bibinfo {year} {2010})}\BibitemShut {NoStop}%
\bibitem [{\citenamefont {Abrosimov}\ \emph {et~al.}(2017)\citenamefont {Abrosimov}, \citenamefont {Aref’Ev}, \citenamefont {Becker}, \citenamefont {Bettin}, \citenamefont {Bulanov}, \citenamefont {Churbanov}, \citenamefont {Filimonov}, \citenamefont {Gavva}, \citenamefont {Godisov}, \citenamefont {Gusev}, \citenamefont {Kotereva}, \citenamefont {Nietzold}, \citenamefont {Peters}, \citenamefont {Potapov}, \citenamefont {Pohl}, \citenamefont {Pramann}, \citenamefont {Riemann}, \citenamefont {Scheel}, \citenamefont {Stosch}, \citenamefont {Wundrack},\ and\ \citenamefont {Zakel}}]{abrosimov2017new}%
  \BibitemOpen
  \bibfield  {author} {\bibinfo {author} {\bibfnamefont {N.~V.}\ \bibnamefont {Abrosimov}}, \bibinfo {author} {\bibfnamefont {D.~G.}\ \bibnamefont {Aref’Ev}}, \bibinfo {author} {\bibfnamefont {P.}~\bibnamefont {Becker}}, \bibinfo {author} {\bibfnamefont {H.}~\bibnamefont {Bettin}}, \bibinfo {author} {\bibfnamefont {A.~D.}\ \bibnamefont {Bulanov}}, \bibinfo {author} {\bibfnamefont {M.~F.}\ \bibnamefont {Churbanov}}, \bibinfo {author} {\bibfnamefont {S.~V.}\ \bibnamefont {Filimonov}}, \bibinfo {author} {\bibfnamefont {V.~A.}\ \bibnamefont {Gavva}}, \bibinfo {author} {\bibfnamefont {O.~N.}\ \bibnamefont {Godisov}}, \bibinfo {author} {\bibfnamefont {A.~V.}\ \bibnamefont {Gusev}}, \bibinfo {author} {\bibfnamefont {T.~V.}\ \bibnamefont {Kotereva}}, \bibinfo {author} {\bibfnamefont {D.}~\bibnamefont {Nietzold}}, \bibinfo {author} {\bibfnamefont {M.}~\bibnamefont {Peters}}, \bibinfo {author} {\bibfnamefont {A.~M.}\ \bibnamefont {Potapov}}, \bibinfo {author} {\bibfnamefont {H.-J.}\ \bibnamefont {Pohl}}, \bibinfo
  {author} {\bibfnamefont {A.}~\bibnamefont {Pramann}}, \bibinfo {author} {\bibfnamefont {H.}~\bibnamefont {Riemann}}, \bibinfo {author} {\bibfnamefont {P.-T.}\ \bibnamefont {Scheel}}, \bibinfo {author} {\bibfnamefont {R.}~\bibnamefont {Stosch}}, \bibinfo {author} {\bibfnamefont {S.}~\bibnamefont {Wundrack}},\ and\ \bibinfo {author} {\bibfnamefont {S.}~\bibnamefont {Zakel}},\ }\bibfield  {title} {\bibinfo {title} {{A new generation of 99.999\% enriched 28Si single crystals for the determination of Avogadro’s constant}},\ }\href@noop {} {\bibfield  {journal} {\bibinfo  {journal} {{Metrologia}}\ }\textbf {\bibinfo {volume} {54}},\ \bibinfo {pages} {599} (\bibinfo {year} {2017})}\BibitemShut {NoStop}%
\bibitem [{\citenamefont {Ager}\ \emph {et~al.}(2005)\citenamefont {Ager}, \citenamefont {Beeman}, \citenamefont {Hansen}, \citenamefont {Haller}, \citenamefont {Sharp}, \citenamefont {Liao}, \citenamefont {Yang}, \citenamefont {Thewalt},\ and\ \citenamefont {Riemann}}]{ager2005high}%
  \BibitemOpen
  \bibfield  {author} {\bibinfo {author} {\bibfnamefont {J.~W.}\ \bibnamefont {Ager}}, \bibinfo {author} {\bibfnamefont {J.~W.}\ \bibnamefont {Beeman}}, \bibinfo {author} {\bibfnamefont {W.~L.}\ \bibnamefont {Hansen}}, \bibinfo {author} {\bibfnamefont {E.~E.}\ \bibnamefont {Haller}}, \bibinfo {author} {\bibfnamefont {I.~D.}\ \bibnamefont {Sharp}}, \bibinfo {author} {\bibfnamefont {C.}~\bibnamefont {Liao}}, \bibinfo {author} {\bibfnamefont {A.}~\bibnamefont {Yang}}, \bibinfo {author} {\bibfnamefont {M.~L.~W.}\ \bibnamefont {Thewalt}},\ and\ \bibinfo {author} {\bibfnamefont {H.}~\bibnamefont {Riemann}},\ }\bibfield  {title} {\bibinfo {title} {{High-purity, isotopically enriched bulk silicon}},\ }\href@noop {} {\bibfield  {journal} {\bibinfo  {journal} {{Journal of the Electrochemical Society}}\ }\textbf {\bibinfo {volume} {152}},\ \bibinfo {pages} {G448} (\bibinfo {year} {2005})}\BibitemShut {NoStop}%
\bibitem [{\citenamefont {Mazzocchi}\ \emph {et~al.}(2019)\citenamefont {Mazzocchi}, \citenamefont {Sennikov}, \citenamefont {Bulanov}, \citenamefont {Churbanov}, \citenamefont {Bertrand}, \citenamefont {Hutin}, \citenamefont {Barnes}, \citenamefont {Drozdov}, \citenamefont {Hartmann},\ and\ \citenamefont {Sanquer}}]{mazzocchi201999}%
  \BibitemOpen
  \bibfield  {author} {\bibinfo {author} {\bibfnamefont {V.}~\bibnamefont {Mazzocchi}}, \bibinfo {author} {\bibfnamefont {P.~G.}\ \bibnamefont {Sennikov}}, \bibinfo {author} {\bibfnamefont {A.~D.}\ \bibnamefont {Bulanov}}, \bibinfo {author} {\bibfnamefont {M.~F.}\ \bibnamefont {Churbanov}}, \bibinfo {author} {\bibfnamefont {B.}~\bibnamefont {Bertrand}}, \bibinfo {author} {\bibfnamefont {L.}~\bibnamefont {Hutin}}, \bibinfo {author} {\bibfnamefont {J.~P.}\ \bibnamefont {Barnes}}, \bibinfo {author} {\bibfnamefont {M.~N.}\ \bibnamefont {Drozdov}}, \bibinfo {author} {\bibfnamefont {J.~M.}\ \bibnamefont {Hartmann}},\ and\ \bibinfo {author} {\bibfnamefont {M.}~\bibnamefont {Sanquer}},\ }\bibfield  {title} {\bibinfo {title} {{99.992\% 28Si CVD-grown epilayer on 300 mm substrates for large scale integration of silicon spin qubits}},\ }\href@noop {} {\bibfield  {journal} {\bibinfo  {journal} {{Journal of Crystal Growth}}\ }\textbf {\bibinfo {volume} {509}},\ \bibinfo {pages} {1} (\bibinfo {year} {2019})}\BibitemShut
  {NoStop}%
\bibitem [{\citenamefont {Liu}\ \emph {et~al.}(2022)\citenamefont {Liu}, \citenamefont {Rinner}, \citenamefont {Remmele}, \citenamefont {Ernst}, \citenamefont {Reiserer},\ and\ \citenamefont {Boeck}}]{liu202228silicon}%
  \BibitemOpen
  \bibfield  {author} {\bibinfo {author} {\bibfnamefont {Y.}~\bibnamefont {Liu}}, \bibinfo {author} {\bibfnamefont {S.}~\bibnamefont {Rinner}}, \bibinfo {author} {\bibfnamefont {T.}~\bibnamefont {Remmele}}, \bibinfo {author} {\bibfnamefont {O.}~\bibnamefont {Ernst}}, \bibinfo {author} {\bibfnamefont {A.}~\bibnamefont {Reiserer}},\ and\ \bibinfo {author} {\bibfnamefont {T.}~\bibnamefont {Boeck}},\ }\bibfield  {title} {\bibinfo {title} {{28Silicon-on-insulator for optically interfaced quantum emitters}},\ }\href@noop {} {\bibfield  {journal} {\bibinfo  {journal} {{Journal of Crystal Growth}}\ }\textbf {\bibinfo {volume} {593}},\ \bibinfo {pages} {126733} (\bibinfo {year} {2022})}\BibitemShut {NoStop}%
\bibitem [{\citenamefont {Itoh}\ \emph {et~al.}(2003)\citenamefont {Itoh}, \citenamefont {Kato}, \citenamefont {Uemura}, \citenamefont {Kaliteevskii}, \citenamefont {Godisov}, \citenamefont {Devyatych}, \citenamefont {Bulanov}, \citenamefont {Gusev}, \citenamefont {Kovalev}, \citenamefont {Sennikov}, \citenamefont {Pohl}, \citenamefont {Abrosimov},\ and\ \citenamefont {Riemann}}]{itoh2003high}%
  \BibitemOpen
  \bibfield  {author} {\bibinfo {author} {\bibfnamefont {K.~M.}\ \bibnamefont {Itoh}}, \bibinfo {author} {\bibfnamefont {J.}~\bibnamefont {Kato}}, \bibinfo {author} {\bibfnamefont {M.}~\bibnamefont {Uemura}}, \bibinfo {author} {\bibfnamefont {A.~K.}\ \bibnamefont {Kaliteevskii}}, \bibinfo {author} {\bibfnamefont {O.~N.}\ \bibnamefont {Godisov}}, \bibinfo {author} {\bibfnamefont {G.~G.}\ \bibnamefont {Devyatych}}, \bibinfo {author} {\bibfnamefont {A.~D.}\ \bibnamefont {Bulanov}}, \bibinfo {author} {\bibfnamefont {A.~V.}\ \bibnamefont {Gusev}}, \bibinfo {author} {\bibfnamefont {I.~D.}\ \bibnamefont {Kovalev}}, \bibinfo {author} {\bibfnamefont {P.~G.}\ \bibnamefont {Sennikov}}, \bibinfo {author} {\bibfnamefont {H.-J.}\ \bibnamefont {Pohl}}, \bibinfo {author} {\bibfnamefont {N.~V.}\ \bibnamefont {Abrosimov}},\ and\ \bibinfo {author} {\bibfnamefont {H.}~\bibnamefont {Riemann}},\ }\bibfield  {title} {\bibinfo {title} {{High purity isotopically enriched 29Si and 30Si single crystals: Isotope separation,
  purification, and growth}},\ }\href@noop {} {\bibfield  {journal} {\bibinfo  {journal} {{Japanese Journal of Applied Physics}}\ }\textbf {\bibinfo {volume} {42}},\ \bibinfo {pages} {6248} (\bibinfo {year} {2003})}\BibitemShut {NoStop}%
\bibitem [{\citenamefont {Devyatykh}\ \emph {et~al.}(2008)\citenamefont {Devyatykh}, \citenamefont {Bulanov}, \citenamefont {Gusev}, \citenamefont {Kovalev}, \citenamefont {Krylov}, \citenamefont {Potapov}, \citenamefont {Sennikov}, \citenamefont {Adamchik}, \citenamefont {Gavva}, \citenamefont {Kotkov}, \citenamefont {Churbanov}, \citenamefont {Dianov}, \citenamefont {Kaliteevskii}, \citenamefont {Godisov}, \citenamefont {Pohl}, \citenamefont {Becker}, \citenamefont {Riemann},\ and\ \citenamefont {Abrosimov}}]{Devyatykh2008}%
  \BibitemOpen
  \bibfield  {author} {\bibinfo {author} {\bibfnamefont {G.~G.}\ \bibnamefont {Devyatykh}}, \bibinfo {author} {\bibfnamefont {A.~D.}\ \bibnamefont {Bulanov}}, \bibinfo {author} {\bibfnamefont {A.~V.}\ \bibnamefont {Gusev}}, \bibinfo {author} {\bibfnamefont {I.~D.}\ \bibnamefont {Kovalev}}, \bibinfo {author} {\bibfnamefont {V.~A.}\ \bibnamefont {Krylov}}, \bibinfo {author} {\bibfnamefont {A.~M.}\ \bibnamefont {Potapov}}, \bibinfo {author} {\bibfnamefont {P.~G.}\ \bibnamefont {Sennikov}}, \bibinfo {author} {\bibfnamefont {S.~A.}\ \bibnamefont {Adamchik}}, \bibinfo {author} {\bibfnamefont {V.~A.}\ \bibnamefont {Gavva}}, \bibinfo {author} {\bibfnamefont {A.~P.}\ \bibnamefont {Kotkov}}, \bibinfo {author} {\bibfnamefont {M.~F.}\ \bibnamefont {Churbanov}}, \bibinfo {author} {\bibfnamefont {E.~M.}\ \bibnamefont {Dianov}}, \bibinfo {author} {\bibfnamefont {A.~K.}\ \bibnamefont {Kaliteevskii}}, \bibinfo {author} {\bibfnamefont {O.~N.}\ \bibnamefont {Godisov}}, \bibinfo {author} {\bibfnamefont {H.~J.}\ \bibnamefont
  {Pohl}}, \bibinfo {author} {\bibfnamefont {P.}~\bibnamefont {Becker}}, \bibinfo {author} {\bibfnamefont {H.}~\bibnamefont {Riemann}},\ and\ \bibinfo {author} {\bibfnamefont {N.~V.}\ \bibnamefont {Abrosimov}},\ }\bibfield  {title} {\bibinfo {title} {{High-purity single-crystal monoisotopic silicon-28 for precise determination of Avogadro's number}},\ }\href {https://doi.org/10.1134/S001250080807001X} {\bibfield  {journal} {\bibinfo  {journal} {{Doklady Chemistry}}\ }\textbf {\bibinfo {volume} {421}},\ \bibinfo {pages} {157} (\bibinfo {year} {2008})}\BibitemShut {NoStop}%
\bibitem [{\citenamefont {Klos}\ \emph {et~al.}(2024)\citenamefont {Klos}, \citenamefont {Tr{\"o}ger}, \citenamefont {Keutgen}, \citenamefont {Losert}, \citenamefont {Abrosimov}, \citenamefont {Knoch}, \citenamefont {Bracht}, \citenamefont {Coppersmith}, \citenamefont {Friesen}, \citenamefont {Cojocaru-Mir{\'e}din}, \citenamefont {Schreiber},\ and\ \citenamefont {Bougeard}}]{klos2024atomistic}%
  \BibitemOpen
  \bibfield  {author} {\bibinfo {author} {\bibfnamefont {J.}~\bibnamefont {Klos}}, \bibinfo {author} {\bibfnamefont {J.}~\bibnamefont {Tr{\"o}ger}}, \bibinfo {author} {\bibfnamefont {J.}~\bibnamefont {Keutgen}}, \bibinfo {author} {\bibfnamefont {M.~P.}\ \bibnamefont {Losert}}, \bibinfo {author} {\bibfnamefont {N.~V.}\ \bibnamefont {Abrosimov}}, \bibinfo {author} {\bibfnamefont {J.}~\bibnamefont {Knoch}}, \bibinfo {author} {\bibfnamefont {H.}~\bibnamefont {Bracht}}, \bibinfo {author} {\bibfnamefont {S.~N.}\ \bibnamefont {Coppersmith}}, \bibinfo {author} {\bibfnamefont {M.}~\bibnamefont {Friesen}}, \bibinfo {author} {\bibfnamefont {O.}~\bibnamefont {Cojocaru-Mir{\'e}din}}, \bibinfo {author} {\bibfnamefont {L.~R.}\ \bibnamefont {Schreiber}},\ and\ \bibinfo {author} {\bibfnamefont {D.}~\bibnamefont {Bougeard}},\ }\bibfield  {title} {\bibinfo {title} {{Atomistic Compositional Details and Their Importance for Spin Qubits in Isotope-Purified Silicon Quantum Wells}},\ }\href@noop {} {\bibfield  {journal} {\bibinfo
  {journal} {{Advanced Science}}\ }\textbf {\bibinfo {volume} {11}},\ \bibinfo {pages} {2407442} (\bibinfo {year} {2024})}\BibitemShut {NoStop}%
\bibitem [{\citenamefont {Becker}(2012)}]{becker2012new}%
  \BibitemOpen
  \bibfield  {author} {\bibinfo {author} {\bibfnamefont {P.}~\bibnamefont {Becker}},\ }\bibfield  {title} {\bibinfo {title} {The new kilogram definition based on counting the atoms in a 28si crystal},\ }\href@noop {} {\bibfield  {journal} {\bibinfo  {journal} {Contemporary Physics}\ }\textbf {\bibinfo {volume} {53}},\ \bibinfo {pages} {461} (\bibinfo {year} {2012})}\BibitemShut {NoStop}%
\bibitem [{\citenamefont {Johnson}\ \emph {et~al.}(2015)\citenamefont {Johnson}, \citenamefont {McCallum},\ and\ \citenamefont {Aziz}}]{johnson2015solid}%
  \BibitemOpen
  \bibfield  {author} {\bibinfo {author} {\bibfnamefont {B.~C.}\ \bibnamefont {Johnson}}, \bibinfo {author} {\bibfnamefont {J.~C.}\ \bibnamefont {McCallum}},\ and\ \bibinfo {author} {\bibfnamefont {M.~J.}\ \bibnamefont {Aziz}},\ }\bibfield  {title} {\bibinfo {title} {{Solid-phase epitaxy}},\ }in\ \href@noop {} {\emph {\bibinfo {booktitle} {{Handbook of Crystal Growth}}}}\ (\bibinfo  {publisher} {{Elsevier}},\ \bibinfo {year} {2015})\ pp.\ \bibinfo {pages} {317--363}\BibitemShut {NoStop}%
\bibitem [{\citenamefont {Radek}\ \emph {et~al.}(2015)\citenamefont {Radek}, \citenamefont {Bracht}, \citenamefont {Johnson}, \citenamefont {McCallum}, \citenamefont {Posselt},\ and\ \citenamefont {Liedke}}]{Radek2015}%
  \BibitemOpen
  \bibfield  {author} {\bibinfo {author} {\bibfnamefont {M.}~\bibnamefont {Radek}}, \bibinfo {author} {\bibfnamefont {H.}~\bibnamefont {Bracht}}, \bibinfo {author} {\bibfnamefont {B.~C.}\ \bibnamefont {Johnson}}, \bibinfo {author} {\bibfnamefont {J.~C.}\ \bibnamefont {McCallum}}, \bibinfo {author} {\bibfnamefont {M.}~\bibnamefont {Posselt}},\ and\ \bibinfo {author} {\bibfnamefont {B.}~\bibnamefont {Liedke}},\ }\bibfield  {title} {\bibinfo {title} {{Atomic transport during solid-phase epitaxial recrystallization of amorphous germanium}},\ }\href {https://doi.org/10.1063/1.4929839} {\bibfield  {journal} {\bibinfo  {journal} {{Applied Physics Letters}}\ }\textbf {\bibinfo {volume} {107}},\ \bibinfo {pages} {082112} (\bibinfo {year} {2015})}\BibitemShut {NoStop}%
\bibitem [{\citenamefont {Bracht}\ \emph {et~al.}(1998)\citenamefont {Bracht}, \citenamefont {Haller},\ and\ \citenamefont {Clark-Phelps}}]{bracht1998silicon}%
  \BibitemOpen
  \bibfield  {author} {\bibinfo {author} {\bibfnamefont {H.}~\bibnamefont {Bracht}}, \bibinfo {author} {\bibfnamefont {E.~E.}\ \bibnamefont {Haller}},\ and\ \bibinfo {author} {\bibfnamefont {R.}~\bibnamefont {Clark-Phelps}},\ }\bibfield  {title} {\bibinfo {title} {{Silicon self-diffusion in isotope heterostructures}},\ }\href@noop {} {\bibfield  {journal} {\bibinfo  {journal} {{Physical Review Letters}}\ }\textbf {\bibinfo {volume} {81}},\ \bibinfo {pages} {393} (\bibinfo {year} {1998})}\BibitemShut {NoStop}%
\bibitem [{\citenamefont {Holland}\ \emph {et~al.}(1996)\citenamefont {Holland}, \citenamefont {Xie}, \citenamefont {Nielsen},\ and\ \citenamefont {Zhou}}]{holland1996implantation}%
  \BibitemOpen
  \bibfield  {author} {\bibinfo {author} {\bibfnamefont {O.~W.}\ \bibnamefont {Holland}}, \bibinfo {author} {\bibfnamefont {L.}~\bibnamefont {Xie}}, \bibinfo {author} {\bibfnamefont {B.}~\bibnamefont {Nielsen}},\ and\ \bibinfo {author} {\bibfnamefont {D.~S.}\ \bibnamefont {Zhou}},\ }\bibfield  {title} {\bibinfo {title} {{Implantation of Si under extreme conditions: the effects of high temperature and dose on damage accumulation}},\ }\href@noop {} {\bibfield  {journal} {\bibinfo  {journal} {{Journal of electronic materials}}\ }\textbf {\bibinfo {volume} {25}},\ \bibinfo {pages} {99} (\bibinfo {year} {1996})}\BibitemShut {NoStop}%
\bibitem [{\citenamefont {Williams}\ \emph {et~al.}(1983)\citenamefont {Williams}, \citenamefont {Chivers}, \citenamefont {Elliman}, \citenamefont {Johnson}, \citenamefont {Lawson}, \citenamefont {Mitchell}, \citenamefont {Orrman-Rossiter}, \citenamefont {Pogany},\ and\ \citenamefont {Short}}]{williams1983production}%
  \BibitemOpen
  \bibfield  {author} {\bibinfo {author} {\bibfnamefont {J.~S.}\ \bibnamefont {Williams}}, \bibinfo {author} {\bibfnamefont {D.~J.}\ \bibnamefont {Chivers}}, \bibinfo {author} {\bibfnamefont {R.~G.}\ \bibnamefont {Elliman}}, \bibinfo {author} {\bibfnamefont {S.~T.}\ \bibnamefont {Johnson}}, \bibinfo {author} {\bibfnamefont {E.~M.}\ \bibnamefont {Lawson}}, \bibinfo {author} {\bibfnamefont {I.~V.}\ \bibnamefont {Mitchell}}, \bibinfo {author} {\bibfnamefont {K.~G.}\ \bibnamefont {Orrman-Rossiter}}, \bibinfo {author} {\bibfnamefont {A.~P.}\ \bibnamefont {Pogany}},\ and\ \bibinfo {author} {\bibfnamefont {K.~T.}\ \bibnamefont {Short}},\ }\bibfield  {title} {\bibinfo {title} {{The Production of Porous Structures on Si, Ge and GaAs by High Dose Ion Implantation}},\ }\href@noop {} {\bibfield  {journal} {\bibinfo  {journal} {{MRS Online Proceedings Library (OPL)}}\ }\textbf {\bibinfo {volume} {27}},\ \bibinfo {pages} {205} (\bibinfo {year} {1983})}\BibitemShut {NoStop}%
\bibitem [{\citenamefont {Williams}(1994)}]{williams1994ion}%
  \BibitemOpen
  \bibfield  {author} {\bibinfo {author} {\bibfnamefont {J.~S.}\ \bibnamefont {Williams}},\ }\bibfield  {title} {\bibinfo {title} {{Ion Induced Damage and Dynamic Annealing Processes}},\ }in\ \href@noop {} {\emph {\bibinfo {booktitle} {{Laser and Ion Beam Modification of Materials}}}}\ (\bibinfo  {publisher} {{Elsevier}},\ \bibinfo {year} {1994})\ pp.\ \bibinfo {pages} {417--423}\BibitemShut {NoStop}%
\bibitem [{\citenamefont {Zhu}\ \emph {et~al.}(1997)\citenamefont {Zhu}, \citenamefont {Williams}, \citenamefont {Llewellyn},\ and\ \citenamefont {McCallum}}]{zhu1997microstructure}%
  \BibitemOpen
  \bibfield  {author} {\bibinfo {author} {\bibfnamefont {X.}~\bibnamefont {Zhu}}, \bibinfo {author} {\bibfnamefont {J.~S.}\ \bibnamefont {Williams}}, \bibinfo {author} {\bibfnamefont {D.~J.}\ \bibnamefont {Llewellyn}},\ and\ \bibinfo {author} {\bibfnamefont {J.~C.}\ \bibnamefont {McCallum}},\ }\bibfield  {title} {\bibinfo {title} {{Microstructure of Ultra High Dose Self Implanted Silicon}},\ }\href@noop {} {\bibfield  {journal} {\bibinfo  {journal} {{MRS Online Proceedings Library}}\ }\textbf {\bibinfo {volume} {504}},\ \bibinfo {pages} {27} (\bibinfo {year} {1997})}\BibitemShut {NoStop}%
\bibitem [{\citenamefont {Goldberg}\ \emph {et~al.}(1999)\citenamefont {Goldberg}, \citenamefont {Williams},\ and\ \citenamefont {Elliman}}]{goldberg1999preferential}%
  \BibitemOpen
  \bibfield  {author} {\bibinfo {author} {\bibfnamefont {R.~D.}\ \bibnamefont {Goldberg}}, \bibinfo {author} {\bibfnamefont {J.~S.}\ \bibnamefont {Williams}},\ and\ \bibinfo {author} {\bibfnamefont {R.~G.}\ \bibnamefont {Elliman}},\ }\bibfield  {title} {\bibinfo {title} {{Preferential amorphization at extended defects of self-ion-irradiated silicon}},\ }\href@noop {} {\bibfield  {journal} {\bibinfo  {journal} {{Physical Review Letters}}\ }\textbf {\bibinfo {volume} {82}},\ \bibinfo {pages} {771} (\bibinfo {year} {1999})}\BibitemShut {NoStop}%
\bibitem [{\citenamefont {Brown}\ \emph {et~al.}(1998)\citenamefont {Brown}, \citenamefont {Kononchuk}, \citenamefont {Rozgonyi}, \citenamefont {Koveshnikov}, \citenamefont {Knights}, \citenamefont {Simpson},\ and\ \citenamefont {Gonzalez}}]{brown1998impurity}%
  \BibitemOpen
  \bibfield  {author} {\bibinfo {author} {\bibfnamefont {R.~A.}\ \bibnamefont {Brown}}, \bibinfo {author} {\bibfnamefont {O.}~\bibnamefont {Kononchuk}}, \bibinfo {author} {\bibfnamefont {G.~A.}\ \bibnamefont {Rozgonyi}}, \bibinfo {author} {\bibfnamefont {S.}~\bibnamefont {Koveshnikov}}, \bibinfo {author} {\bibfnamefont {A.~P.}\ \bibnamefont {Knights}}, \bibinfo {author} {\bibfnamefont {P.~J.}\ \bibnamefont {Simpson}},\ and\ \bibinfo {author} {\bibfnamefont {F.}~\bibnamefont {Gonzalez}},\ }\bibfield  {title} {\bibinfo {title} {{Impurity gettering to secondary defects created by MeV ion implantation in silicon}},\ }\href@noop {} {\bibfield  {journal} {\bibinfo  {journal} {{Journal of Applied Physics}}\ }\textbf {\bibinfo {volume} {84}},\ \bibinfo {pages} {2459} (\bibinfo {year} {1998})}\BibitemShut {NoStop}%
\bibitem [{\citenamefont {Venezia}\ \emph {et~al.}(1998)\citenamefont {Venezia}, \citenamefont {Eaglesham}, \citenamefont {Haynes}, \citenamefont {Agarwal}, \citenamefont {Jacobson}, \citenamefont {Gossmann},\ and\ \citenamefont {Baumann}}]{venezia1998depth}%
  \BibitemOpen
  \bibfield  {author} {\bibinfo {author} {\bibfnamefont {V.~C.}\ \bibnamefont {Venezia}}, \bibinfo {author} {\bibfnamefont {D.~J.}\ \bibnamefont {Eaglesham}}, \bibinfo {author} {\bibfnamefont {T.~E.}\ \bibnamefont {Haynes}}, \bibinfo {author} {\bibfnamefont {A.}~\bibnamefont {Agarwal}}, \bibinfo {author} {\bibfnamefont {D.~C.}\ \bibnamefont {Jacobson}}, \bibinfo {author} {\bibfnamefont {H.-J.}\ \bibnamefont {Gossmann}},\ and\ \bibinfo {author} {\bibfnamefont {F.~H.}\ \bibnamefont {Baumann}},\ }\bibfield  {title} {\bibinfo {title} {{Depth profiling of vacancy clusters in MeV-implanted Si using Au labeling}},\ }\href@noop {} {\bibfield  {journal} {\bibinfo  {journal} {{Applied Physics Letters}}\ }\textbf {\bibinfo {volume} {73}},\ \bibinfo {pages} {2980} (\bibinfo {year} {1998})}\BibitemShut {NoStop}%
\bibitem [{\citenamefont {Williams}\ \emph {et~al.}(2001)\citenamefont {Williams}, \citenamefont {Conway}, \citenamefont {Williams},\ and\ \citenamefont {Wong-Leung}}]{williams2001direct}%
  \BibitemOpen
  \bibfield  {author} {\bibinfo {author} {\bibfnamefont {J.~S.}\ \bibnamefont {Williams}}, \bibinfo {author} {\bibfnamefont {M.~J.}\ \bibnamefont {Conway}}, \bibinfo {author} {\bibfnamefont {B.~C.}\ \bibnamefont {Williams}},\ and\ \bibinfo {author} {\bibfnamefont {J.}~\bibnamefont {Wong-Leung}},\ }\bibfield  {title} {\bibinfo {title} {{Direct observation of voids in the vacancy excess region of ion bombarded silicon}},\ }\href@noop {} {\bibfield  {journal} {\bibinfo  {journal} {{Applied Physics Letters}}\ }\textbf {\bibinfo {volume} {78}},\ \bibinfo {pages} {2867} (\bibinfo {year} {2001})}\BibitemShut {NoStop}%
\bibitem [{\citenamefont {Peeva}\ \emph {et~al.}(2000)\citenamefont {Peeva}, \citenamefont {Koegler}, \citenamefont {Brauer}, \citenamefont {Werner},\ and\ \citenamefont {Skorupa}}]{peeva2000metallic}%
  \BibitemOpen
  \bibfield  {author} {\bibinfo {author} {\bibfnamefont {A.}~\bibnamefont {Peeva}}, \bibinfo {author} {\bibfnamefont {R.}~\bibnamefont {Koegler}}, \bibinfo {author} {\bibfnamefont {G.}~\bibnamefont {Brauer}}, \bibinfo {author} {\bibfnamefont {P.}~\bibnamefont {Werner}},\ and\ \bibinfo {author} {\bibfnamefont {W.}~\bibnamefont {Skorupa}},\ }\bibfield  {title} {\bibinfo {title} {{Metallic impurity gettering to defects remaining in the RP/2 region of MeV-ion implanted and annealed silicon}},\ }\href@noop {} {\bibfield  {journal} {\bibinfo  {journal} {{Materials Science in Semiconductor Processing}}\ }\textbf {\bibinfo {volume} {3}},\ \bibinfo {pages} {297} (\bibinfo {year} {2000})}\BibitemShut {NoStop}%
\bibitem [{\citenamefont {Williams}\ and\ \citenamefont {Wong-Leung}(2009)}]{williams2009voids}%
  \BibitemOpen
  \bibfield  {author} {\bibinfo {author} {\bibfnamefont {J.~S.}\ \bibnamefont {Williams}}\ and\ \bibinfo {author} {\bibfnamefont {J.}~\bibnamefont {Wong-Leung}},\ }\bibfield  {title} {\bibinfo {title} {{Voids and nanocavities in silicon}},\ }\href@noop {} {\bibfield  {journal} {\bibinfo  {journal} {{Materials Science with Ion Beams}}\ ,\ \bibinfo {pages} {113}} (\bibinfo {year} {2009})}\BibitemShut {NoStop}%
\bibitem [{\citenamefont {Mazzone}(1986)}]{mazzone1986defect}%
  \BibitemOpen
  \bibfield  {author} {\bibinfo {author} {\bibfnamefont {A.~M.}\ \bibnamefont {Mazzone}},\ }\bibfield  {title} {\bibinfo {title} {{Defect distribution in ion-implanted silicon}},\ }\href@noop {} {\bibfield  {journal} {\bibinfo  {journal} {{Physica Status Solidi. A, Applied Research}}\ }\textbf {\bibinfo {volume} {95}},\ \bibinfo {pages} {149} (\bibinfo {year} {1986})}\BibitemShut {NoStop}%
\bibitem [{\citenamefont {Pellegrino}\ \emph {et~al.}(2001)\citenamefont {Pellegrino}, \citenamefont {Leveque}, \citenamefont {Wong-Leung}, \citenamefont {Jagadish},\ and\ \citenamefont {Svensson}}]{pellegrino2001separation}%
  \BibitemOpen
  \bibfield  {author} {\bibinfo {author} {\bibfnamefont {P.}~\bibnamefont {Pellegrino}}, \bibinfo {author} {\bibfnamefont {P.}~\bibnamefont {Leveque}}, \bibinfo {author} {\bibfnamefont {J.}~\bibnamefont {Wong-Leung}}, \bibinfo {author} {\bibfnamefont {C.}~\bibnamefont {Jagadish}},\ and\ \bibinfo {author} {\bibfnamefont {B.~G.}\ \bibnamefont {Svensson}},\ }\bibfield  {title} {\bibinfo {title} {{Separation of vacancy and interstitial depth profiles in ion-implanted silicon: Experimental observation}},\ }\href@noop {} {\bibfield  {journal} {\bibinfo  {journal} {{Applied Physics Letters}}\ }\textbf {\bibinfo {volume} {78}},\ \bibinfo {pages} {3442} (\bibinfo {year} {2001})}\BibitemShut {NoStop}%
\bibitem [{\citenamefont {Kudriavtsev}\ \emph {et~al.}(2020)\citenamefont {Kudriavtsev}, \citenamefont {Hernandez-Zanabria}, \citenamefont {Salinas},\ and\ \citenamefont {Asomoza}}]{kudriavtsev2020formation}%
  \BibitemOpen
  \bibfield  {author} {\bibinfo {author} {\bibfnamefont {Y.}~\bibnamefont {Kudriavtsev}}, \bibinfo {author} {\bibfnamefont {A.}~\bibnamefont {Hernandez-Zanabria}}, \bibinfo {author} {\bibfnamefont {C.}~\bibnamefont {Salinas}},\ and\ \bibinfo {author} {\bibfnamefont {R.}~\bibnamefont {Asomoza}},\ }\bibfield  {title} {\bibinfo {title} {{The formation of porous silicon by irradiation with low-energy ions}},\ }\href@noop {} {\bibfield  {journal} {\bibinfo  {journal} {{Vacuum}}\ }\textbf {\bibinfo {volume} {177}},\ \bibinfo {pages} {109393} (\bibinfo {year} {2020})}\BibitemShut {NoStop}%
\bibitem [{\citenamefont {Zhu}\ \emph {et~al.}(1999)\citenamefont {Zhu}, \citenamefont {Williams}, \citenamefont {Llewellyn},\ and\ \citenamefont {McCallum}}]{zhu1999instability}%
  \BibitemOpen
  \bibfield  {author} {\bibinfo {author} {\bibfnamefont {X.}~\bibnamefont {Zhu}}, \bibinfo {author} {\bibfnamefont {J.~S.}\ \bibnamefont {Williams}}, \bibinfo {author} {\bibfnamefont {D.~J.}\ \bibnamefont {Llewellyn}},\ and\ \bibinfo {author} {\bibfnamefont {J.~C.}\ \bibnamefont {McCallum}},\ }\bibfield  {title} {\bibinfo {title} {{Instability of nanocavities in amorphous silicon}},\ }\href@noop {} {\bibfield  {journal} {\bibinfo  {journal} {{Applied Physics Letters}}\ }\textbf {\bibinfo {volume} {74}},\ \bibinfo {pages} {2313} (\bibinfo {year} {1999})}\BibitemShut {NoStop}%
\bibitem [{\citenamefont {Zhu}\ \emph {et~al.}(2001)\citenamefont {Zhu}, \citenamefont {Williams}, \citenamefont {Conway}, \citenamefont {Ridgway}, \citenamefont {Fortuna}, \citenamefont {Ruault},\ and\ \citenamefont {Bernas}}]{zhu2001direct}%
  \BibitemOpen
  \bibfield  {author} {\bibinfo {author} {\bibfnamefont {X.~F.}\ \bibnamefont {Zhu}}, \bibinfo {author} {\bibfnamefont {J.~S.}\ \bibnamefont {Williams}}, \bibinfo {author} {\bibfnamefont {M.~J.}\ \bibnamefont {Conway}}, \bibinfo {author} {\bibfnamefont {M.~C.}\ \bibnamefont {Ridgway}}, \bibinfo {author} {\bibfnamefont {F.}~\bibnamefont {Fortuna}}, \bibinfo {author} {\bibfnamefont {M.-O.}\ \bibnamefont {Ruault}},\ and\ \bibinfo {author} {\bibfnamefont {H.}~\bibnamefont {Bernas}},\ }\bibfield  {title} {\bibinfo {title} {{Direct observation of irradiation-induced nanocavity shrinkage in Si}},\ }\href@noop {} {\bibfield  {journal} {\bibinfo  {journal} {{Applied Physics Letters}}\ }\textbf {\bibinfo {volume} {79}},\ \bibinfo {pages} {3416} (\bibinfo {year} {2001})}\BibitemShut {NoStop}%
\bibitem [{\citenamefont {Aspelmeyer}\ \emph {et~al.}(2014)\citenamefont {Aspelmeyer}, \citenamefont {Kippenberg},\ and\ \citenamefont {Marquardt}}]{aspelmeyer2014cavity}%
  \BibitemOpen
  \bibfield  {author} {\bibinfo {author} {\bibfnamefont {M.}~\bibnamefont {Aspelmeyer}}, \bibinfo {author} {\bibfnamefont {T.~J.}\ \bibnamefont {Kippenberg}},\ and\ \bibinfo {author} {\bibfnamefont {F.}~\bibnamefont {Marquardt}},\ }\bibfield  {title} {\bibinfo {title} {{Cavity optomechanics}},\ }\href@noop {} {\bibfield  {journal} {\bibinfo  {journal} {{Reviews of Modern Physics}}\ }\textbf {\bibinfo {volume} {86}},\ \bibinfo {pages} {1391} (\bibinfo {year} {2014})}\BibitemShut {NoStop}%
\bibitem [{\citenamefont {Kumar}\ \emph {et~al.}(2022)\citenamefont {Kumar}, \citenamefont {N{\"a}tkinniemi}, \citenamefont {Lyyra},\ and\ \citenamefont {Muhonen}}]{kumar2022single}%
  \BibitemOpen
  \bibfield  {author} {\bibinfo {author} {\bibfnamefont {A.~S.}\ \bibnamefont {Kumar}}, \bibinfo {author} {\bibfnamefont {J.}~\bibnamefont {N{\"a}tkinniemi}}, \bibinfo {author} {\bibfnamefont {H.}~\bibnamefont {Lyyra}},\ and\ \bibinfo {author} {\bibfnamefont {J.~T.}\ \bibnamefont {Muhonen}},\ }\bibfield  {title} {\bibinfo {title} {{Single-laser feedback cooling of optomechanical resonators}},\ }\href@noop {} {\bibfield  {journal} {\bibinfo  {journal} {{arXiv preprint arXiv:2209.06029}}\ } (\bibinfo {year} {2022})}\BibitemShut {NoStop}%
\bibitem [{\citenamefont {Doolittle}(1985)}]{doolittle1985algorithms}%
  \BibitemOpen
  \bibfield  {author} {\bibinfo {author} {\bibfnamefont {L.~R.}\ \bibnamefont {Doolittle}},\ }\bibfield  {title} {\bibinfo {title} {{Algorithms for the rapid simulation of Rutherford backscattering spectra}},\ }\href@noop {} {\bibfield  {journal} {\bibinfo  {journal} {{Nuclear Instruments and Methods in Physics Research Section B: Beam Interactions with Materials and Atoms}}\ }\textbf {\bibinfo {volume} {9}},\ \bibinfo {pages} {344} (\bibinfo {year} {1985})}\BibitemShut {NoStop}%
\bibitem [{\citenamefont {Ziegler}\ \emph {et~al.}(2008)\citenamefont {Ziegler}, \citenamefont {Biersack}, \citenamefont {Ziegler} \emph {et~al.}}]{ziegler2008stopping}%
  \BibitemOpen
  \bibfield  {author} {\bibinfo {author} {\bibfnamefont {J.~F.}\ \bibnamefont {Ziegler}}, \bibinfo {author} {\bibfnamefont {J.~P.}\ \bibnamefont {Biersack}}, \bibinfo {author} {\bibfnamefont {M.~D.}\ \bibnamefont {Ziegler}}, \emph {et~al.},\ }\href@noop {} {\emph {\bibinfo {title} {{The stopping and range of ions in matter}}}}\ (\bibinfo  {publisher} {{SRIM Chester, Maryland}},\ \bibinfo {year} {2008})\BibitemShut {NoStop}%
\bibitem [{\citenamefont {M{\"o}ller}\ \emph {et~al.}(1988)\citenamefont {M{\"o}ller}, \citenamefont {Eckstein},\ and\ \citenamefont {Biersack}}]{moller1988tridyn}%
  \BibitemOpen
  \bibfield  {author} {\bibinfo {author} {\bibfnamefont {W.}~\bibnamefont {M{\"o}ller}}, \bibinfo {author} {\bibfnamefont {W.}~\bibnamefont {Eckstein}},\ and\ \bibinfo {author} {\bibfnamefont {J.~P.}\ \bibnamefont {Biersack}},\ }\bibfield  {title} {\bibinfo {title} {{Tridyn-binary collision simulation of atomic collisions and dynamic composition changes in solids}},\ }\href@noop {} {\bibfield  {journal} {\bibinfo  {journal} {{Computer Physics Communications}}\ }\textbf {\bibinfo {volume} {51}},\ \bibinfo {pages} {355} (\bibinfo {year} {1988})}\BibitemShut {NoStop}%
\bibitem [{\citenamefont {Custer}\ \emph {et~al.}(1994)\citenamefont {Custer}, \citenamefont {Thompson}, \citenamefont {Jacobson}, \citenamefont {Poate}, \citenamefont {Roorda}, \citenamefont {Sinke},\ and\ \citenamefont {Spaepen}}]{custer1994density}%
  \BibitemOpen
  \bibfield  {author} {\bibinfo {author} {\bibfnamefont {J.~S.}\ \bibnamefont {Custer}}, \bibinfo {author} {\bibfnamefont {M.~O.}\ \bibnamefont {Thompson}}, \bibinfo {author} {\bibfnamefont {D.}~\bibnamefont {Jacobson}}, \bibinfo {author} {\bibfnamefont {J.~M.}\ \bibnamefont {Poate}}, \bibinfo {author} {\bibfnamefont {S.}~\bibnamefont {Roorda}}, \bibinfo {author} {\bibfnamefont {W.~C.}\ \bibnamefont {Sinke}},\ and\ \bibinfo {author} {\bibfnamefont {F.}~\bibnamefont {Spaepen}},\ }\bibfield  {title} {\bibinfo {title} {{Density of amorphous Si}},\ }\href@noop {} {\bibfield  {journal} {\bibinfo  {journal} {{Applied Physics Letters}}\ }\textbf {\bibinfo {volume} {64}},\ \bibinfo {pages} {437} (\bibinfo {year} {1994})}\BibitemShut {NoStop}%
\bibitem [{Note1()}]{Note1}%
  \BibitemOpen
  \bibinfo {note} {Derived assuming a Si atomic density of $5\times 10^{22}$ atoms/cm$^3$ and when the sputter yield is zero, $\Delta z=200$ nm for an ion fluence of $10^{18}$ cm$^{-2}$.}\BibitemShut {Stop}%
\bibitem [{\citenamefont {Woo}\ \emph {et~al.}(2016)\citenamefont {Woo}, \citenamefont {Bertoni}, \citenamefont {Choi}, \citenamefont {Nam}, \citenamefont {Castellanos}, \citenamefont {Powell}, \citenamefont {Buonassisi},\ and\ \citenamefont {Choi}}]{woo2016insight}%
  \BibitemOpen
  \bibfield  {author} {\bibinfo {author} {\bibfnamefont {S.}~\bibnamefont {Woo}}, \bibinfo {author} {\bibfnamefont {M.}~\bibnamefont {Bertoni}}, \bibinfo {author} {\bibfnamefont {K.}~\bibnamefont {Choi}}, \bibinfo {author} {\bibfnamefont {S.}~\bibnamefont {Nam}}, \bibinfo {author} {\bibfnamefont {S.}~\bibnamefont {Castellanos}}, \bibinfo {author} {\bibfnamefont {D.~M.}\ \bibnamefont {Powell}}, \bibinfo {author} {\bibfnamefont {T.}~\bibnamefont {Buonassisi}},\ and\ \bibinfo {author} {\bibfnamefont {H.}~\bibnamefont {Choi}},\ }\bibfield  {title} {\bibinfo {title} {{An insight into dislocation density reduction in multicrystalline silicon}},\ }\href@noop {} {\bibfield  {journal} {\bibinfo  {journal} {{Solar Energy Materials and Solar Cells}}\ }\textbf {\bibinfo {volume} {155}},\ \bibinfo {pages} {88} (\bibinfo {year} {2016})}\BibitemShut {NoStop}%
\bibitem [{\citenamefont {Haunschild}\ \emph {et~al.}(2010)\citenamefont {Haunschild}, \citenamefont {Glatthaar}, \citenamefont {Demant}, \citenamefont {Nievendick}, \citenamefont {Motzko}, \citenamefont {Rein},\ and\ \citenamefont {Weber}}]{haunschild2010quality}%
  \BibitemOpen
  \bibfield  {author} {\bibinfo {author} {\bibfnamefont {J.}~\bibnamefont {Haunschild}}, \bibinfo {author} {\bibfnamefont {M.}~\bibnamefont {Glatthaar}}, \bibinfo {author} {\bibfnamefont {M.}~\bibnamefont {Demant}}, \bibinfo {author} {\bibfnamefont {J.}~\bibnamefont {Nievendick}}, \bibinfo {author} {\bibfnamefont {M.}~\bibnamefont {Motzko}}, \bibinfo {author} {\bibfnamefont {S.}~\bibnamefont {Rein}},\ and\ \bibinfo {author} {\bibfnamefont {E.~R.}\ \bibnamefont {Weber}},\ }\bibfield  {title} {\bibinfo {title} {{Quality control of as-cut multicrystalline silicon wafers using photoluminescence imaging for solar cell production}},\ }\href@noop {} {\bibfield  {journal} {\bibinfo  {journal} {{Solar Energy Materials and Solar Cells}}\ }\textbf {\bibinfo {volume} {94}},\ \bibinfo {pages} {2007} (\bibinfo {year} {2010})}\BibitemShut {NoStop}%
\bibitem [{\citenamefont {Arguirov}\ \emph {et~al.}(2009)\citenamefont {Arguirov}, \citenamefont {Mchedlidze}, \citenamefont {Kittler}, \citenamefont {Reiche}, \citenamefont {Wilhelm}, \citenamefont {Hoang}, \citenamefont {Holleman},\ and\ \citenamefont {Schmitz}}]{arguirov2009silicon}%
  \BibitemOpen
  \bibfield  {author} {\bibinfo {author} {\bibfnamefont {T.}~\bibnamefont {Arguirov}}, \bibinfo {author} {\bibfnamefont {T.}~\bibnamefont {Mchedlidze}}, \bibinfo {author} {\bibfnamefont {M.}~\bibnamefont {Kittler}}, \bibinfo {author} {\bibfnamefont {M.}~\bibnamefont {Reiche}}, \bibinfo {author} {\bibfnamefont {T.}~\bibnamefont {Wilhelm}}, \bibinfo {author} {\bibfnamefont {T.}~\bibnamefont {Hoang}}, \bibinfo {author} {\bibfnamefont {J.}~\bibnamefont {Holleman}},\ and\ \bibinfo {author} {\bibfnamefont {J.}~\bibnamefont {Schmitz}},\ }\bibfield  {title} {\bibinfo {title} {{Silicon based light emitters utilizing radiation from dislocations; electric field induced shift of the dislocation-related luminescence}},\ }\href@noop {} {\bibfield  {journal} {\bibinfo  {journal} {{Physica E: Low-dimensional Systems and Nanostructures}}\ }\textbf {\bibinfo {volume} {41}},\ \bibinfo {pages} {907} (\bibinfo {year} {2009})}\BibitemShut {NoStop}%
\bibitem [{\citenamefont {Stowe}\ \emph {et~al.}(2003)\citenamefont {Stowe}, \citenamefont {Galloway}, \citenamefont {Senkader}, \citenamefont {Mallik}, \citenamefont {Falster},\ and\ \citenamefont {Wilshaw}}]{stowe2003near}%
  \BibitemOpen
  \bibfield  {author} {\bibinfo {author} {\bibfnamefont {D.~J.}\ \bibnamefont {Stowe}}, \bibinfo {author} {\bibfnamefont {S.~A.}\ \bibnamefont {Galloway}}, \bibinfo {author} {\bibfnamefont {S.}~\bibnamefont {Senkader}}, \bibinfo {author} {\bibfnamefont {K.}~\bibnamefont {Mallik}}, \bibinfo {author} {\bibfnamefont {R.~J.}\ \bibnamefont {Falster}},\ and\ \bibinfo {author} {\bibfnamefont {P.~R.}\ \bibnamefont {Wilshaw}},\ }\bibfield  {title} {\bibinfo {title} {{Near-band gap luminescence at room temperature from dislocations in silicon}},\ }\href@noop {} {\bibfield  {journal} {\bibinfo  {journal} {{Physica B: Condensed Matter}}\ }\textbf {\bibinfo {volume} {340}},\ \bibinfo {pages} {710} (\bibinfo {year} {2003})}\BibitemShut {NoStop}%
\bibitem [{\citenamefont {Sigmund}(1969)}]{sigmund1969theory}%
  \BibitemOpen
  \bibfield  {author} {\bibinfo {author} {\bibfnamefont {P.}~\bibnamefont {Sigmund}},\ }\bibfield  {title} {\bibinfo {title} {{Theory of sputtering. I. Sputtering yield of amorphous and polycrystalline targets}},\ }\href@noop {} {\bibfield  {journal} {\bibinfo  {journal} {{Physical Review}}\ }\textbf {\bibinfo {volume} {184}},\ \bibinfo {pages} {383} (\bibinfo {year} {1969})}\BibitemShut {NoStop}%
\bibitem [{\citenamefont {Alm{\'e}n}\ and\ \citenamefont {Bruce}(1961)}]{almen1961sputtering}%
  \BibitemOpen
  \bibfield  {author} {\bibinfo {author} {\bibfnamefont {O.}~\bibnamefont {Alm{\'e}n}}\ and\ \bibinfo {author} {\bibfnamefont {G.}~\bibnamefont {Bruce}},\ }\bibfield  {title} {\bibinfo {title} {{Sputtering experiments in the high energy region}},\ }\href@noop {} {\bibfield  {journal} {\bibinfo  {journal} {{Nuclear Instruments and Methods}}\ }\textbf {\bibinfo {volume} {11}},\ \bibinfo {pages} {279} (\bibinfo {year} {1961})}\BibitemShut {NoStop}%
\bibitem [{\citenamefont {Stevie}(2005)}]{stevie2005focused}%
  \BibitemOpen
  \bibfield  {author} {\bibinfo {author} {\bibfnamefont {F.~A.}\ \bibnamefont {Stevie}},\ }\bibfield  {title} {\bibinfo {title} {{Focused ion beam secondary ion mass spectrometry (FIB-SIMS)}},\ }\href@noop {} {\bibfield  {journal} {\bibinfo  {journal} {{Introduction to Focused Ion Beams: Instrumentation, Theory, Techniques and Practice}}\ ,\ \bibinfo {pages} {269}} (\bibinfo {year} {2005})}\BibitemShut {NoStop}%
\bibitem [{\citenamefont {Jacobson}\ \emph {et~al.}(1986)\citenamefont {Jacobson}, \citenamefont {Poate},\ and\ \citenamefont {Olson}}]{jacobson1986zone}%
  \BibitemOpen
  \bibfield  {author} {\bibinfo {author} {\bibfnamefont {D.~C.}\ \bibnamefont {Jacobson}}, \bibinfo {author} {\bibfnamefont {J.~M.}\ \bibnamefont {Poate}},\ and\ \bibinfo {author} {\bibfnamefont {G.~L.}\ \bibnamefont {Olson}},\ }\bibfield  {title} {\bibinfo {title} {{Zone refining and enhancement of solid phase epitaxial growth rates in Au-implanted amorphous Si}},\ }\href@noop {} {\bibfield  {journal} {\bibinfo  {journal} {{Applied Physics Letters}}\ }\textbf {\bibinfo {volume} {48}},\ \bibinfo {pages} {118} (\bibinfo {year} {1986})}\BibitemShut {NoStop}%
\bibitem [{\citenamefont {G{\"o}sele}\ \emph {et~al.}(1980)\citenamefont {G{\"o}sele}, \citenamefont {Frank},\ and\ \citenamefont {Seeger}}]{gosele1980mechanism}%
  \BibitemOpen
  \bibfield  {author} {\bibinfo {author} {\bibfnamefont {U.}~\bibnamefont {G{\"o}sele}}, \bibinfo {author} {\bibfnamefont {W.}~\bibnamefont {Frank}},\ and\ \bibinfo {author} {\bibfnamefont {A.}~\bibnamefont {Seeger}},\ }\bibfield  {title} {\bibinfo {title} {Mechanism and kinetics of the diffusion of gold in silicon},\ }\href@noop {} {\bibfield  {journal} {\bibinfo  {journal} {{Applied Physics}}\ }\textbf {\bibinfo {volume} {23}},\ \bibinfo {pages} {361} (\bibinfo {year} {1980})}\BibitemShut {NoStop}%
\bibitem [{\citenamefont {Wong-Leung}\ \emph {et~al.}(1995)\citenamefont {Wong-Leung}, \citenamefont {Nygren},\ and\ \citenamefont {Williams}}]{wong1995gettering}%
  \BibitemOpen
  \bibfield  {author} {\bibinfo {author} {\bibfnamefont {J.}~\bibnamefont {Wong-Leung}}, \bibinfo {author} {\bibfnamefont {E.}~\bibnamefont {Nygren}},\ and\ \bibinfo {author} {\bibfnamefont {J.~S.}\ \bibnamefont {Williams}},\ }\bibfield  {title} {\bibinfo {title} {{Gettering of Au to dislocations and cavities in silicon}},\ }\href@noop {} {\bibfield  {journal} {\bibinfo  {journal} {{Applied Physics Letters}}\ }\textbf {\bibinfo {volume} {67}},\ \bibinfo {pages} {416} (\bibinfo {year} {1995})}\BibitemShut {NoStop}%
\bibitem [{\citenamefont {Roth}\ \emph {et~al.}(1990)\citenamefont {Roth}, \citenamefont {Olson}, \citenamefont {Jacobson},\ and\ \citenamefont {Poate}}]{roth1990kinetics}%
  \BibitemOpen
  \bibfield  {author} {\bibinfo {author} {\bibfnamefont {J.~A.}\ \bibnamefont {Roth}}, \bibinfo {author} {\bibfnamefont {G.~L.}\ \bibnamefont {Olson}}, \bibinfo {author} {\bibfnamefont {D.~C.}\ \bibnamefont {Jacobson}},\ and\ \bibinfo {author} {\bibfnamefont {J.~M.}\ \bibnamefont {Poate}},\ }\bibfield  {title} {\bibinfo {title} {{Kinetics of solid phase epitaxy in thick amorphous Si layers formed by MeV ion implantation}},\ }\href@noop {} {\bibfield  {journal} {\bibinfo  {journal} {{Applied Physics Letters}}\ }\textbf {\bibinfo {volume} {57}},\ \bibinfo {pages} {1340} (\bibinfo {year} {1990})}\BibitemShut {NoStop}%
\bibitem [{\citenamefont {Frank}\ \emph {et~al.}(1996)\citenamefont {Frank}, \citenamefont {Gustin},\ and\ \citenamefont {Horz}}]{frank1996diffusion}%
  \BibitemOpen
  \bibfield  {author} {\bibinfo {author} {\bibfnamefont {W.}~\bibnamefont {Frank}}, \bibinfo {author} {\bibfnamefont {W.}~\bibnamefont {Gustin}},\ and\ \bibinfo {author} {\bibfnamefont {M.}~\bibnamefont {Horz}},\ }\bibfield  {title} {\bibinfo {title} {{Diffusion of gold and platinum in amorphous silicon and germanium}},\ }\href@noop {} {\bibfield  {journal} {\bibinfo  {journal} {{Journal of non-crystalline solids}}\ }\textbf {\bibinfo {volume} {205}},\ \bibinfo {pages} {208} (\bibinfo {year} {1996})}\BibitemShut {NoStop}%
\bibitem [{\citenamefont {Elliman}\ \emph {et~al.}(1985)\citenamefont {Elliman}, \citenamefont {Gibson}, \citenamefont {Jacobson}, \citenamefont {Poate},\ and\ \citenamefont {Williams}}]{elliman1985diffusion}%
  \BibitemOpen
  \bibfield  {author} {\bibinfo {author} {\bibfnamefont {R.~G.}\ \bibnamefont {Elliman}}, \bibinfo {author} {\bibfnamefont {J.~M.}\ \bibnamefont {Gibson}}, \bibinfo {author} {\bibfnamefont {D.~C.}\ \bibnamefont {Jacobson}}, \bibinfo {author} {\bibfnamefont {J.~M.}\ \bibnamefont {Poate}},\ and\ \bibinfo {author} {\bibfnamefont {J.~S.}\ \bibnamefont {Williams}},\ }\bibfield  {title} {\bibinfo {title} {{Diffusion and precipitation in amorphous Si}},\ }\href@noop {} {\bibfield  {journal} {\bibinfo  {journal} {{Applied Physics Letters}}\ }\textbf {\bibinfo {volume} {46}},\ \bibinfo {pages} {478} (\bibinfo {year} {1985})}\BibitemShut {NoStop}%
\bibitem [{\citenamefont {Baumann}\ and\ \citenamefont {Schr{\"o}ter}(1991)}]{baumann1991precipitation}%
  \BibitemOpen
  \bibfield  {author} {\bibinfo {author} {\bibfnamefont {F.~H.}\ \bibnamefont {Baumann}}\ and\ \bibinfo {author} {\bibfnamefont {W.}~\bibnamefont {Schr{\"o}ter}},\ }\bibfield  {title} {\bibinfo {title} {Precipitation of gold into metastable gold silicide in silicon},\ }\href@noop {} {\bibfield  {journal} {\bibinfo  {journal} {{Physical Review B}}\ }\textbf {\bibinfo {volume} {43}},\ \bibinfo {pages} {6510} (\bibinfo {year} {1991})}\BibitemShut {NoStop}%
\bibitem [{\citenamefont {Giese}\ \emph {et~al.}(2000)\citenamefont {Giese}, \citenamefont {Bracht}, \citenamefont {Stolwijk},\ and\ \citenamefont {Baither}}]{giese2000microscopic}%
  \BibitemOpen
  \bibfield  {author} {\bibinfo {author} {\bibfnamefont {A.}~\bibnamefont {Giese}}, \bibinfo {author} {\bibfnamefont {H.}~\bibnamefont {Bracht}}, \bibinfo {author} {\bibfnamefont {N.}~\bibnamefont {Stolwijk}},\ and\ \bibinfo {author} {\bibfnamefont {D.}~\bibnamefont {Baither}},\ }\bibfield  {title} {\bibinfo {title} {Microscopic defects in silicon induced by zinc out-diffusion},\ }\href@noop {} {\bibfield  {journal} {\bibinfo  {journal} {{Materials Science and Engineering: B}}\ }\textbf {\bibinfo {volume} {71}},\ \bibinfo {pages} {160} (\bibinfo {year} {2000})}\BibitemShut {NoStop}%
\end{thebibliography}
\end{document}